# The role of terminal groups in non-chiral rod-like compounds on the formation of polar fluids

Michał Czerwiński*, Mateusz Mrukiewicz, Mateusz Filipow, Damian Pociecha, Natalia Podoliak, Dalibor Repček, Monika Zając, Dorota Węgłowska

Michał Czerwiński, Mateusz Filipow, Monika Zając, Dorota Węgłowska
*Institute of Chemistry, Military University of Technology, Kaliskiego 2, 00-908 Warsaw, Poland*

Mateusz Mrukiewicz
*Institute of Applied Physics, Military University of Technology, Kaliskiego 2, 00-908 Warsaw, Poland*

Damian Pociecha
*Faculty of Chemistry, University of Warsaw, Żwirki i Wigury 101, 02-089 Warsaw, Poland*

Natalia Podoliak, Dalibor Repček
*Institute of Physics, Academy of Science of the Czech Republic, Na Slovance 2, 182 00 Prague 8, Czech Republic*

Dalibor Repček
*Faculty of Nuclear Sciences and Physical Engineering, Czech Technical University in Prague, Břehová 7, 110 00 Prague 1, Czech Republic*

Corresponding author: michal.czerwinski@wat.edu.pl

## Abstract

The emergence of ferroelectric mesophases in non-chiral liquid crystal (LCs) has sparked fundamental interest in the molecular mechanisms governing polarity. In this study, we investigate how terminal molecular groups influence the formation and stability of polar phases by analyzing six compounds from three homologous series. Specifically, we compare newly synthesized homologs with a nitro group, which predominantly exhibit polar mesophases, to previously reported structurally related analogs containing either a cyano group or a fluorine atom as terminal fragment. Density Functional Theory (DFT) calculations provide insights into electronic surface potential (ESP) distributions, revealing alternating regions of positive and negative charge density along the molecular axis, consistent with Madhusudana's model of polar phase stabilization. We propose the ESP-derived parameter quantifying terminal electrostatic charge, revealing a direct correlation between the negative-to-positive charge ratio at the molecular termini and the formation of ferroelectric or antiferroelectric mesophases. To validate this hypothesis, we analyze the molecular structure-mesomorphic behavior relationship of other known non-chiral compounds that exhibit polar phases, demonstrating the critical role of terminal groups in determining mesophase polarity. Our findings enhance the understanding of the molecular origins of ferroelectricity in non-chiral LCs, paving the way for the rational design of next-generation functional polar soft materials.

## Introduction

The discovery of polar phases and ferroelectricity in non-chiral liquid crystalline (LC) compounds has generated significant interest in this class of soft materials [1–5]. This growing fascination is driven by the unique combination of fluidity, anisotropic physical properties, and intrinsic ferroelectricity, which closely resembles that observed in solid crystals, particularly in the nematic ferroelectric ($N_F$) phase. In the $N_F$ phase, rod-like molecules align parallel to one another, leading to a coherent orientation of their dipole moment vectors, $\vec{\mu}$, which, when summed, result in a spontaneous polarization vector, $\vec{P}_S$ (Figure 1a). In contrast, in the paraelectric nematic (N) phase, the molecules arrange randomly, causing their dipole moments to cancel out (Figure 1b), thereby not allowing for the spontaneous polarization $\vec{P}_S$ formation.

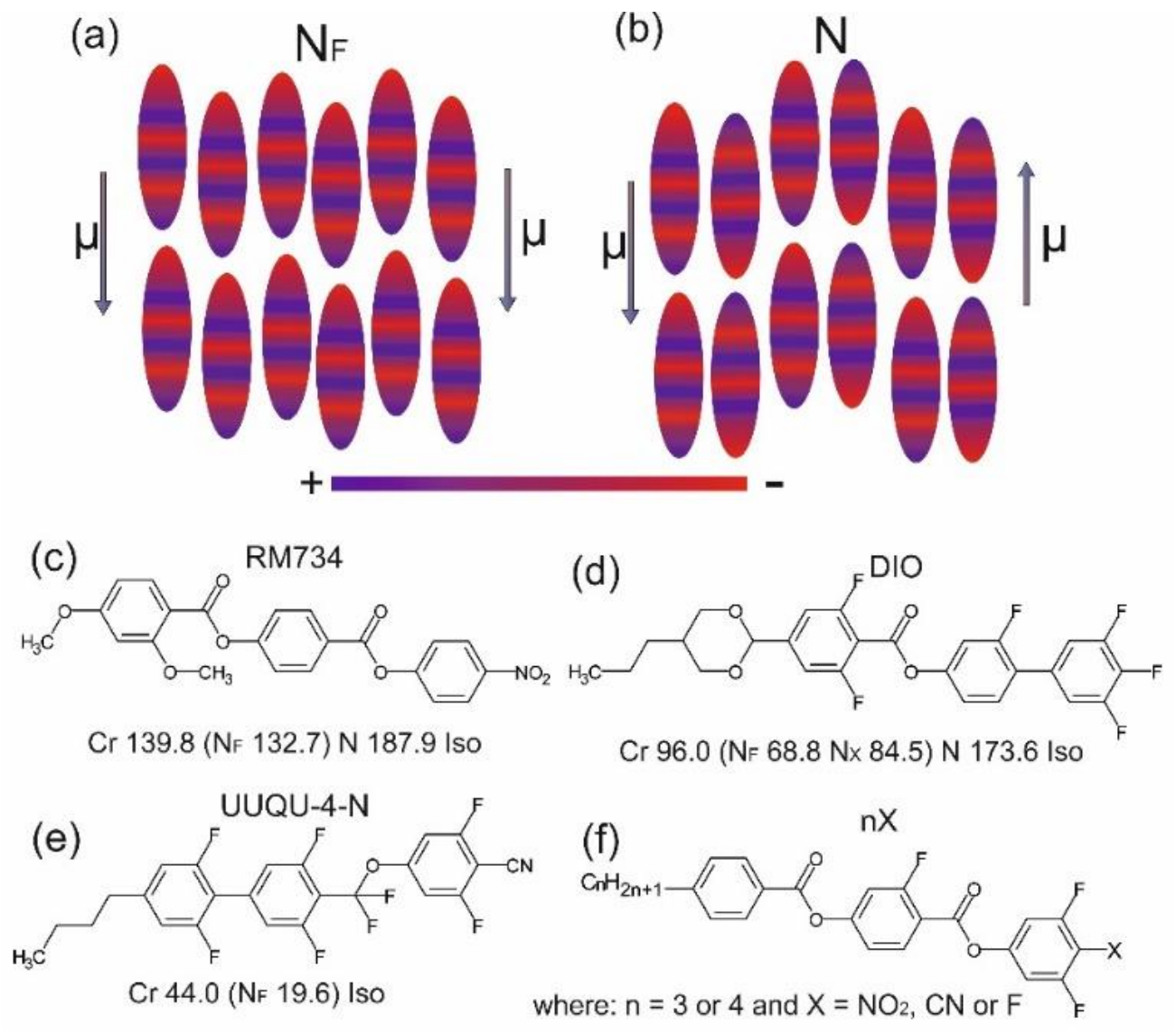

**Figure 1.** A schematic illustration of ordering in a ferroelectric nematic ($N_F$) phase (a) and a paraelectric nematic (N) phase (b) induced by rod-like molecules with altering surface positive (+) and negative (-) charge density across the long axis. Chemical structure of (c) RM734, (d) DIO, and (e) UUQU-4-N compounds with phase transition temperatures; (f) chemical structure of compounds belonging to nX analogs.

Since the $N_F$ phase was first identified in 2017 in the RM734[6] and DIO[7] compounds and, four years later, in UUQU-4-N[8] (Figure 1c, d, and e, respectively), researchers have synthesized numerous new materials exhibiting this phase, drawing inspiration from these archetypal structures.[9–20] Furthermore, an additional nematic LC phase exhibiting microscopic polarity and local antiferroelectric order, known as the $N_x$ phase (also referred to as $SmZ_A$),[21–23] has been discovered. Several polar smectic phases, including $SmA_F$,[24–26] $SmC_F$,[27,28] and $SmC_P$,[29] have also been identified. Additionally, heliconical structures like $N_{TBF}$[4] or $^{HC}N_F$,[30] and $SmC_P^H$[31] or $SmC_{TBF}$[32] have been found, further expanding the landscape of polar LC phases. Beyond these fundamental discoveries, growing attention is being devoted to their potential applications, particularly in non-linear optics,[33,34] electro-optical devices with fast [35] and exceptionally low or near-zero threshold voltage,[1,12,36–38] tunable lasers,[39] piezoelectric elements,[40] and other advanced technologies.[41]

The remarkable potential of non-chiral LC materials with polar phases—both for applications in optoelectronics and photonics and for probing fundamental scientific phenomena—highlights the critical need to develop new compounds and mixtures with these phases.[42–44]Achieving this goal through rational molecular design requires a comprehensive model that accurately describes the formation and stabilization of polar phases. Such a model would facilitate the derivation of structural principles that promote polarity in non-chiral LC systems, thereby unlocking new opportunities for next-generation technologies. The most current and frequently cited model about the emergence of polar phases in non-chiral LCs is the one proposed by Madhusudana.[45] This model approximates polar LC molecules as cylindrical entities with an alternating surface charge density distribution along their long axis. It postulates that electrostatic interactions between oppositely charged regions drive a low-energy parallel configuration, in which neighboring molecules adopt a laterally shifted arrangement (Figure 1a). Recently, Osipov [46,47] demonstrated that Madhusudana's model appears to be more realistic than the one based on electrostatic interactions between permanent longitudinal molecular dipoles. However, he also emphasized the challenge of rigorously validating this model, given that parallel molecular alignment reduces entropy, thereby increasing the system's free energy. To establish the viability of Madhusudana's model, it is essential to demonstrate that the electrostatic interactions it describes lower the system's free energy sufficiently to counteract the entropic cost associated with molecular ordering. As Osipov has shown, a numerical evaluation of this effect remains a formidable challenge and has yet to be undertaken.

To date, only a limited number of experimental studies have directly tested Madhusudana's model by examining how variations in surface charge density along the molecular axis influence LC polarity.[48–50] These studies primarily explore the effects of molecular architecture and fluorine incorporation into the rigid core. It has been observed that an increase in the number of oscillations in the averaged electronic surface potential with a sinusoidal profile favors the emergence and stabilization of polar phases, whereas significant amplitude variations in these oscillations tend to promote either the nematic $N_X$ phase or polar smectic phases. While a few studies have considered the role of terminal groups in non-chiral compounds on the stability of their polar phases,[43,51–53] they typically examine the effect of a single polar terminal group and do not explicitly account for the combined influence of both molecular termini within the framework of Madhusudana's model.

In this study, to conduct an in-depth analysis of the influence of terminal groups on the polarity of LC mesophases, we synthesized and examined two homologues with a nitro terminal group ($nNO_2$), having a rigid core composed of three partially fluorinated benzene rings connected by ester linkages and an alkyl terminal chain of varying length (see general structure in Figure 1f). The obtained results, combined with data for previously studied analogs featuring either a polar terminal cyano group, nCN,[17] or terminal fluorine atom, nF,[54] enabled a systematic evaluation of the impact of both terminal molecular fragments on the stability of polar phases within the framework of Madhusudana's model. Based on this analysis, we introduce a novel parameter and propose an assumption that refines the structural guidelines for designing new non-chiral polar LC materials. These findings provide deeper insight into the chemical structure–property relationships that govern the molecular origins of the polar phase, advancing our understanding of ferro- and antiferroelectricity in non-chiral LC systems.

## Results

The route of synthesis, the synthesis details, and the chemical characterization of compounds $3NO_2$ and $4NO_2$, are described in the ESI.

Following the structural verification of the synthesized $3NO_2$ and $4NO_2$ homologues, a comprehensive study of their mesomorphic properties was conducted. The phase transition temperatures, enthalpy changes, and corresponding scaled entropy values ($\Delta S/R$) were determined and summarized in Table S1 (ESI). This dataset also includes previously synthesized and characterized analogs, nCN[17] and nF.[54] Despite only small variation in terminal alkyl chain length a notable difference in mesomorphic behavior was observed between two synthesized $nNO_2$ homologs. $3NO_2$ exhibits both a melting and clearing point approximately 10 °C higher than $4NO_2$. Furthermore, $3NO_2$ shows a direct phase transition from the ferroelectric nematic phase to the isotropic phase, whereas $4NO_2$ exhibits an enantiotropic sequence involving $N_x$ and N phases, along with a monotropic $N_F$ phase. The presence and nature of the mesophases were confirmed through polarized-ligth optical microscopy (POM) imaging (Figure 2a). The ferroelectric nematic phase in $3NO_2$ is characterized by a multidomain marble-like texture displaying two distinct colors. During the cooling cycle of $4NO_2$, a nematic phase marble texture emerges, transitioning into a highly defected $N_x$ phase texture, which subsequently evolves into a multidomain banded texture characteristic of the $N_F$ phase.

The scaled entropy change ($\Delta S/R$), a measure of disorder alteration across phase transitions, was compared for the synthesized $nNO_2$ compounds and their analogs (nF and nCN). The transition from the paraelectric nematic (N) phase to the isotropic (Iso) phase exhibits typical $\Delta S/R$ values (of the order of tenths) for all compounds listed in Table S1 (ESI). In contrast, the $N_X$–N phase transition is associated with an exceptionally low entropy change, ranging from a few to several thousandths, observed in $4NO_2$ as well as in 3CN and 4CN. These values align with those reported in the literature for this phase transition.[22] A transition between $N_F$ and $N_X$ phases, in contrast, is marked by an order-of-magnitude higher value of $\Delta S/R$, comparable to that observed for the N–Iso transition. This suggests a similar degree of molecular reorganization across the transition from ferroelectric to paraelectric nematic phase and from paraelectric nematic to isotropic phase. Notably, the direct $N_F$–Iso transition in $4NO_2$ exhibits a scaled entropy change an order of magnitude greater than the conventional N–Iso transition, comparable to that observed for the $SmA_F$–N transition in 3F. This pronounced difference likely reflects an additional entropic contribution arising from dipole ordering in the $N_F$ or $SmA_F$ phase, as opposed to the non-polar nature of the conventional nematic phase.

To further investigate the distinct nature of the nematic phases in the studied compounds, a binary mixture system comprising $3NO_2$ and $4NO_2$ was prepared and analyzed (Figure 1b). Enantiotropic ferroelectric nematic phases persist in mixtures containing up to a 0.8 mole fraction of compound $4NO_2$. The $N_x$ phase emerges at and above the equimolar composition of $3NO_2$ and $4NO_2$ mixtures. Overall, the phase transition sequence and corresponding temperatures in the $4NO_2$–$3NO_2$ mixture system corroborate the mesomorphic properties of the studied homologs.

The three-dimensional electronic surface potential (3D-ESP) calculated on optimized geometry using the B3LYP/6-311G+(d,p) level of density functional theory (DFT) reveals an alternating distribution of positive and negative potential regions along the long molecular axis of both $3NO_2$ and $4NO_2$ (Figure 1c). According to the previously described Madhusudana model, this electrostatic potential distribution may explain the predominant occurrence of polar phases in these homologs.

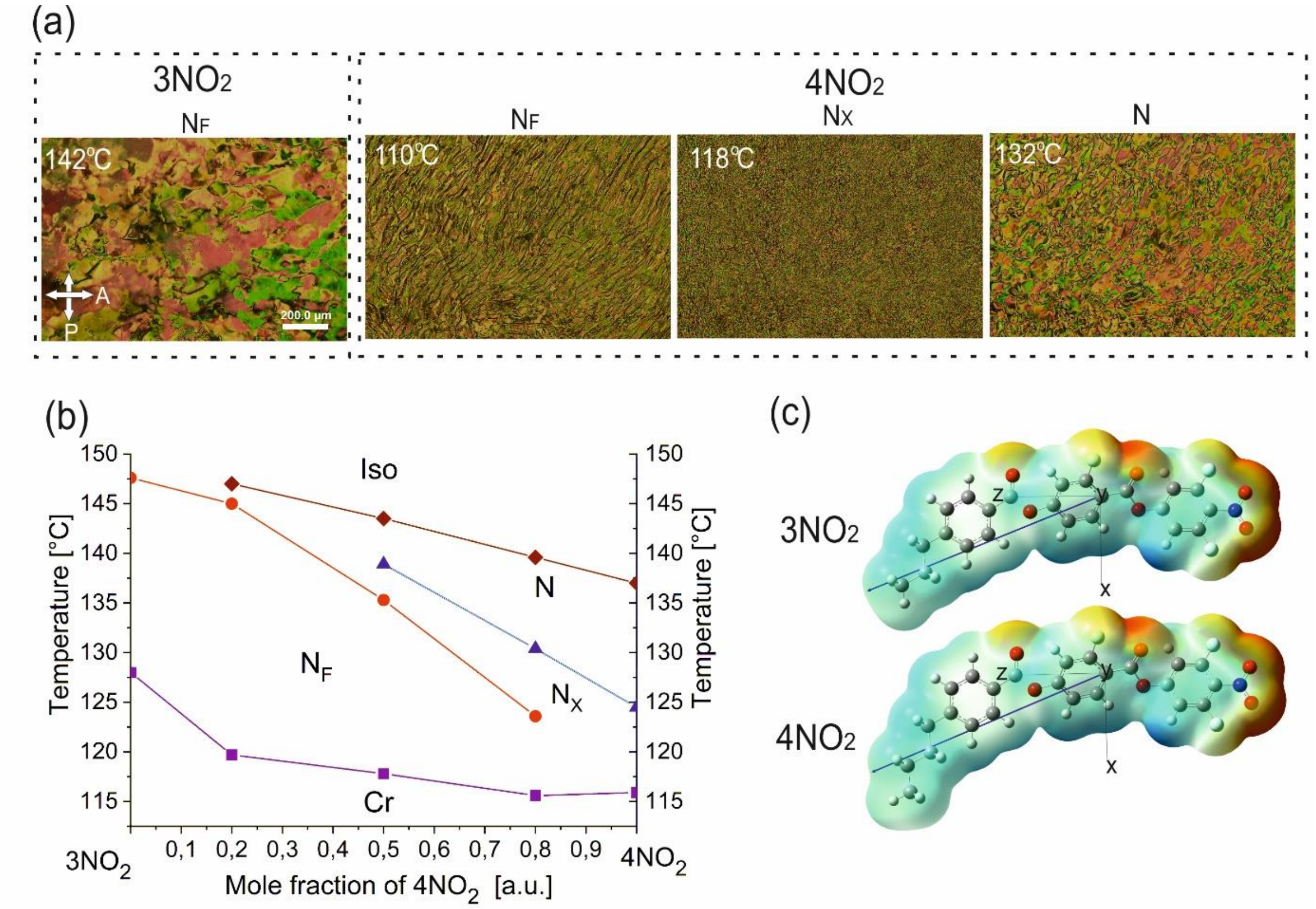

**Figure 2.** Mespomorphic and electrostatic properties of $3NO_2$ and $4NO_2$ homologs, and their mixtures. (a) POM images, obtained for material between untreated glass plates on cooling from the isotropic phase under crossed polarizes. (b) Phase diagram obtained on heating for bicomponent mixtures system $4NO_2$-$3NO_2$. (c) Optimized geometries of $3NO_2$ and $4NO_2$ at B3LYP/6-311G+(d,p) level of DFT, with the electrostatic potential (ESP) on the 0.0004 au electron density iso-surface and marked vector of the dipole moment (blue arrow) with cartesian axes.

To confirm the ferroelectric properties and determine the molecular structure of the newly synthesized $3NO_2$ and $4NO_2$ compounds, we performed complementary studies. Dielectric spectroscopy was done in the cells with uncoated electrodes to avoid the effect of polyimide alignment layers on the measured values of the permittivity.[55–59] Figures 3a and 3b present the real part of electric permittivity ($\varepsilon'$) of $3NO_2$ and $4NO_2$, respectively. In both cases, the transition to the ferroelectric nematic phase ($N_F$) is characterized by a strong increase in measured ε′ value. A similar behavior was observed in 3CN as well in 4CN.[17] On the other hand, the transition to the crystalline phase (Cr) is manifested by a sharp decrease in electrical permittivity. In $3NO_2$ the $N_F$ phase is created directly from the isotropic phase (Iso), therefore the transition appears as the continuous increase of electric permittivity from Iso. However, in the $4NO_2$ compound, the ferroelectric nematic phase is formed from the antiferroelectric nematic phase ($N_x$). In the polar nematics, the N-$N_x$ transition is characterized as a sudden decrease of ε′ in the cooling cycle.[17,60] From the point of view of molecular dynamics, in the ferroelectric nematic phase, we observed a strong collective relaxation inside the ferroelectric domains with a relaxation frequency of a few kHz.[17] In the compound $4NO_2$, the collective mode starts to become visible in the $N_x$ phase together with the molecular mode, as in the case of the results of the dielectric studies performed for 4CN.

The ferroelectric behavior of $3NO_2$ and $4NO_2$ was confirmed by detecting the polarization reversal current peak in the $N_F$ phase, after applying the triangular wave-shaped electric field signal (Figure 3c, d). The current response was obtained with different frequencies of the electric field, which indicate different dynamics of the polarization switching mechanism, and for various resistor resistance values. The complete switching of the spontaneous polarization vector for compound $3NO_2$ occurs in a shorter time than for $4NO_2$. The change in the parameters at which the polarization peak was visible is due to the fact that the $N_F$ phase in the $3NO_2$ compound is formed directly from the isotropic phase, whereas in the $4NO_2$ compound the ferroelectric phase is formed from the $N_x$ phase. The polarization current measurement was performed so that the single current peak in the $N_F$ phase was sufficiently saturated. This allowed us to obtain the spontaneous polarization value at the level of $P_S \sim 0.8$ $\mu C/cm^2$ and $\sim 2.5$ $\mu C/cm^2$ for $3NO_2$ and $4NO_2$, respectively. The $P_S$ value calculated for $4NO_2$ is typical for ferroelectric nematics, which do not show a direct transition from the Iso phase to the $N_F$ phase.[1,61,62]

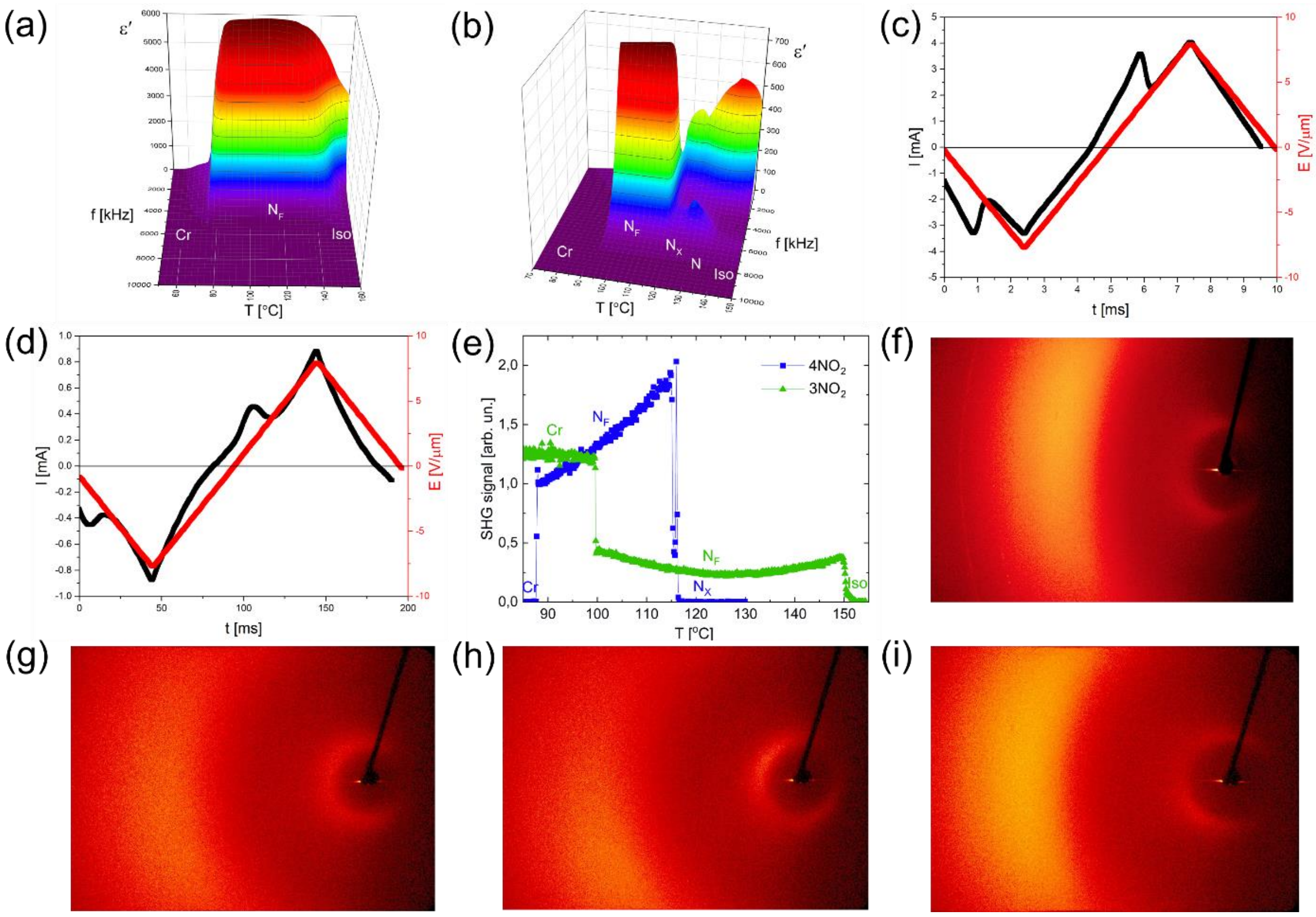

**Figure 3.** Physical properties of $3NO_2$ and $4NO_2$. The real part of electric permittivity (ε′) versus frequency (f) and temperature (T) measured in (a) the $3NO_2$ compound and (b) the $4NO_2$ compound. The current response of the ferroelectric nematic phase in (c) $3NO_2$ at 110°C and (d) $4NO_2$ at 114°C under the triangular wave-shaped electric signal. The results were obtained for different frequencies of the applied field and various resistor resistance values. (e) Temperature dependence of the SHG signal. Two-dimensional XRD patterns were obtained in (f) the $N_F$ phase of $3NO_2$, and (g) the N phase, (h) the $N_X$ phase, (i) the $N_F$ phase of $4NO_2$.

To confirm polar character of $N_F$ phase, SHG measurements were conducted for both $3NO_2$ and $4NO_2$ compounds. The results obtained on cooling from Iso phase are presented in Figure 2e. For $3NO_2$, a sharp increase in SGH intensity signal is observed at the Iso-$N_F$ transition temperature. The intensity of the signal first gradually decreases on cooling within the $N_F$ phase and then after reaching a local minimum started slowly increasing when approaching Cr phase. Remarkably, the crystalline phase of this compound also reveals polar properties. The SHG signal intensity sharply raised at the $N_F$-Cr transition temperature, several times exceeding the values observed in the $N_F$ phase. For $4NO_2$ compound measured on cooling from the Iso phase, no SHG signal was observed in N and $N_x$ phases confirming the absence of macroscopic polar order. At the $N_x$ - $N_F$ transition, an abrupt increase with a subsequent drop down and following rapid increase to approximately the same values is observed, remaining a sharp tooth. Similar behavior was observed for the previously studied 3CN and 4CN compounds,[17] which was explained to occur as a result of pretransitional instabilities. Within the temperature range of the $N_F$ phase, the SHG signal considerably decreases with the temperature lowering. At the $N_F$-Cr phase transition, another sharp “tooth” is observed, followed by the intensity decrease down to zero values, evidencing non-polar crystalline phase for $4NO_2$ compound.

Liquid character of all the liquid crystalline phases formed by $nNO_2$ homologues has been confirmed by X-ray diffraction (XRD). As expected for nematic phases only broad, diffused diffraction signals has been recorded, in both low- and high-angle ranges, evidencing lack of long-range positional order of molecules (Figure 3f-i and Figure S1 and S2). Interestingly, for $4NO_2$ a considerable change in local periodicity along the director has been observed between paraelaectric and ferroelectric nematic phases, from ~23.5 Å to ~20 Å, respectively. Moreover, strong change in preferred molecular orientation took place at the transition to the ferroelectric phase. In one-surface-free (droplet) sample in N phase director aligned perpendicular to the surface sample, while in the $N_F$ phase, exhibiting spontaneous electric polarization, a parallel orientation of director with respect to sample surface has been detected (Figure S3). Such an anchoring transition is driven by tendency to avoid the charges at the film surface.[63]

Sequence of three nematic phases in $4NO_2$ compound was also confirmed by measuring the optical birefringence (Figure S4). As observed previously for other compounds with similar phase sequence transition between N and $N_x$ phases causes only small variation in the slope of $\Delta n(T)$ dependence, which points to a similar degree of molecular order in these phases. The transition to $N_F$ phase is accompanied by considerable increase in birefringence, due to growth of orientational order caused by creation of long-range order of molecular dipoles. However, prior the transition a small deviation of $\Delta n$ is detected, most probably to strong pretransitional fluctuations of director (splay deformation).

**Discussion**

**Phase polarity-terminal groups relationship in $nNO_2$, nCN, and nF analogs**

For comprehensive analysis of the structure-property relationship in the newly synthesized $nNO_2$ compounds and their previously obtained analogs, collective plots illustrating their mesomorphic behavior (Figure 4) and one-dimensional electrostatic potential (1D-ESP) distributions (Figure 5) have been constructed. The 1D-ESP profiles represent the radially averaged electrostatic potential along the long molecular axis at an electron density isovalue of 0.0004. The methodology for generating these plots follows the approach described in other studies [48,49] and is detailed in the experimental section of the electronic supplementary information (ESI). Complete 3D-ESP surfaces and the reduced 1D-ESP plots for all analogs are shown in Figure S5 (ESI).

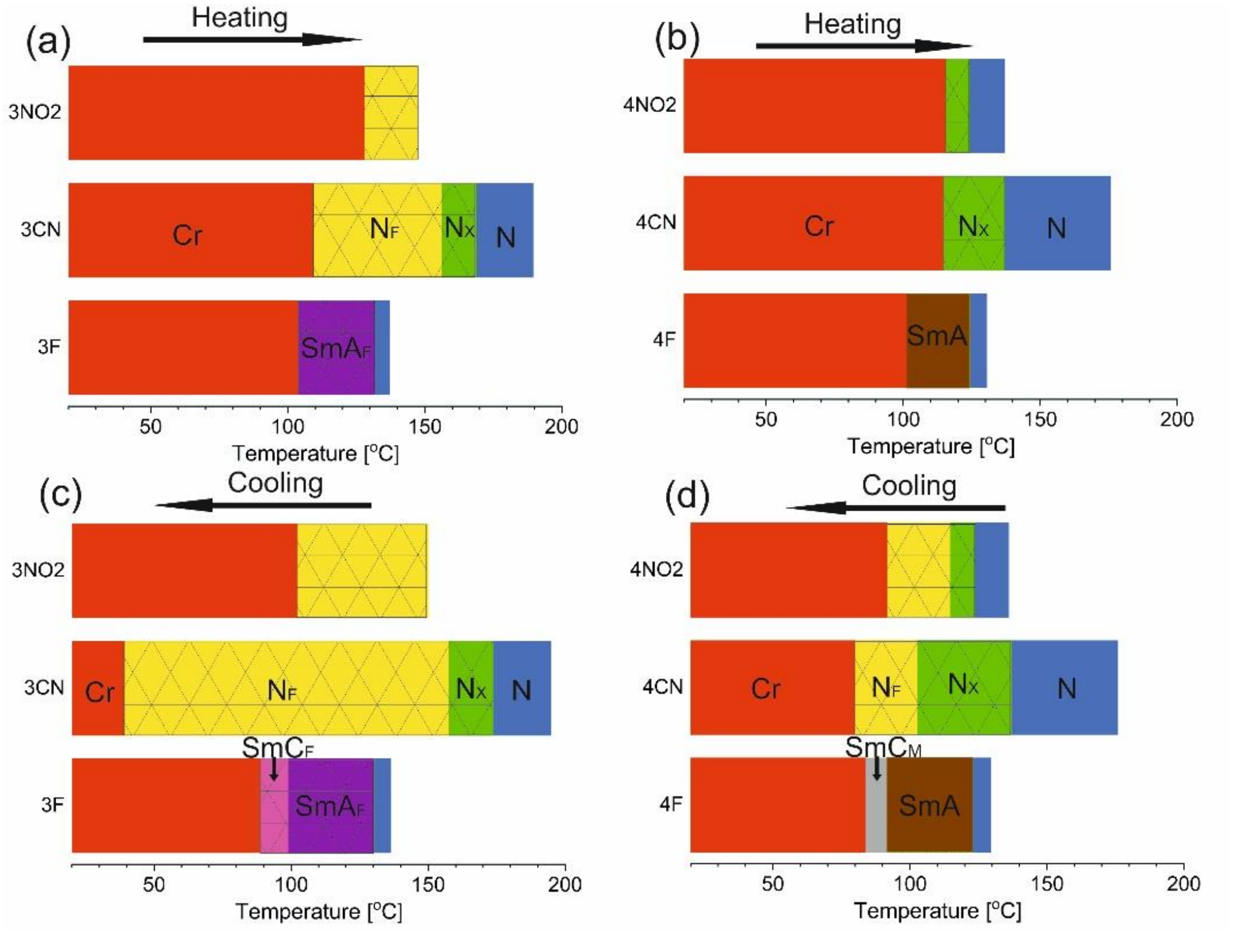

**Figure 4.** Temperature ranges of phases in $nNO_2$, nCN, and nF analogs as a function of heating (a, b) and cooling (c, d). Phases with macroscopic ($N_F$) and local polarity ($N_X$) are indicated by a triangular mesh.

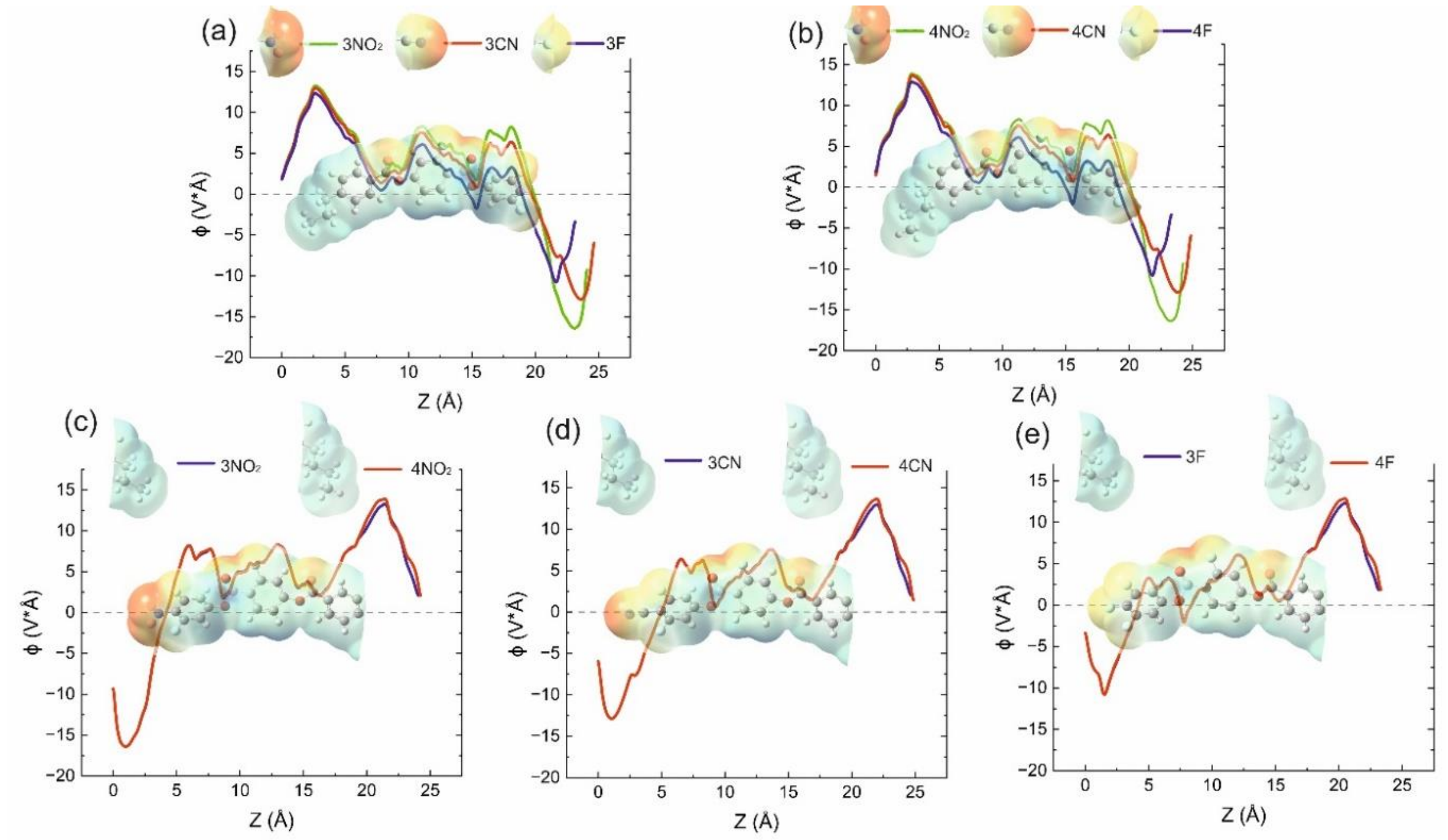

**Figure 5.** Averaged electrostatic potential, scaled by contour length (discussed in the ESI), along the z-axis of the molecule at an electron density isosurface of 0.0004 calculated at the B3LYP-6-311G+(d,p) level of DFT for analogs with different polar terminal group (a-b) and homologs with different terminal alkyl length (c-e).

Furthermore, Table 1 presents key quantitative parameters, including dipole moment values, relative temperature ranges of polar, on a macro- or microscopic level, versus non-polar mesophases, and newly defined parameters $\Psi_{(-)}$ and $\Psi_{(+)}$, which provide a quantitative measure of electrostatic charge distribution within the molecular framework. Specifically, $\Psi_{(-)}$ and $\Psi_{(+)}$ represent the total negative and positive electrostatic charges, respectively, localized within the isocontour surfaces of a given terminal molecular fragment. These parameters are defined by equation (1):

$$\Psi = \int_{z_1}^{z_2} \phi dz \qquad (1)$$

where $\phi$ represents the electric flux, and z denotes the distance along the long molecular axis. Figure 6 illustrates this concept using compound $3NO_2$ as an example, showing the integration limits applied and highlighting the surface areas corresponding to $\Psi_{(-)}$ and $\Psi_{(+)}$. A detailed description of these parameters, along with their derivation from 1D-ESP plots, is provided in the ESI.

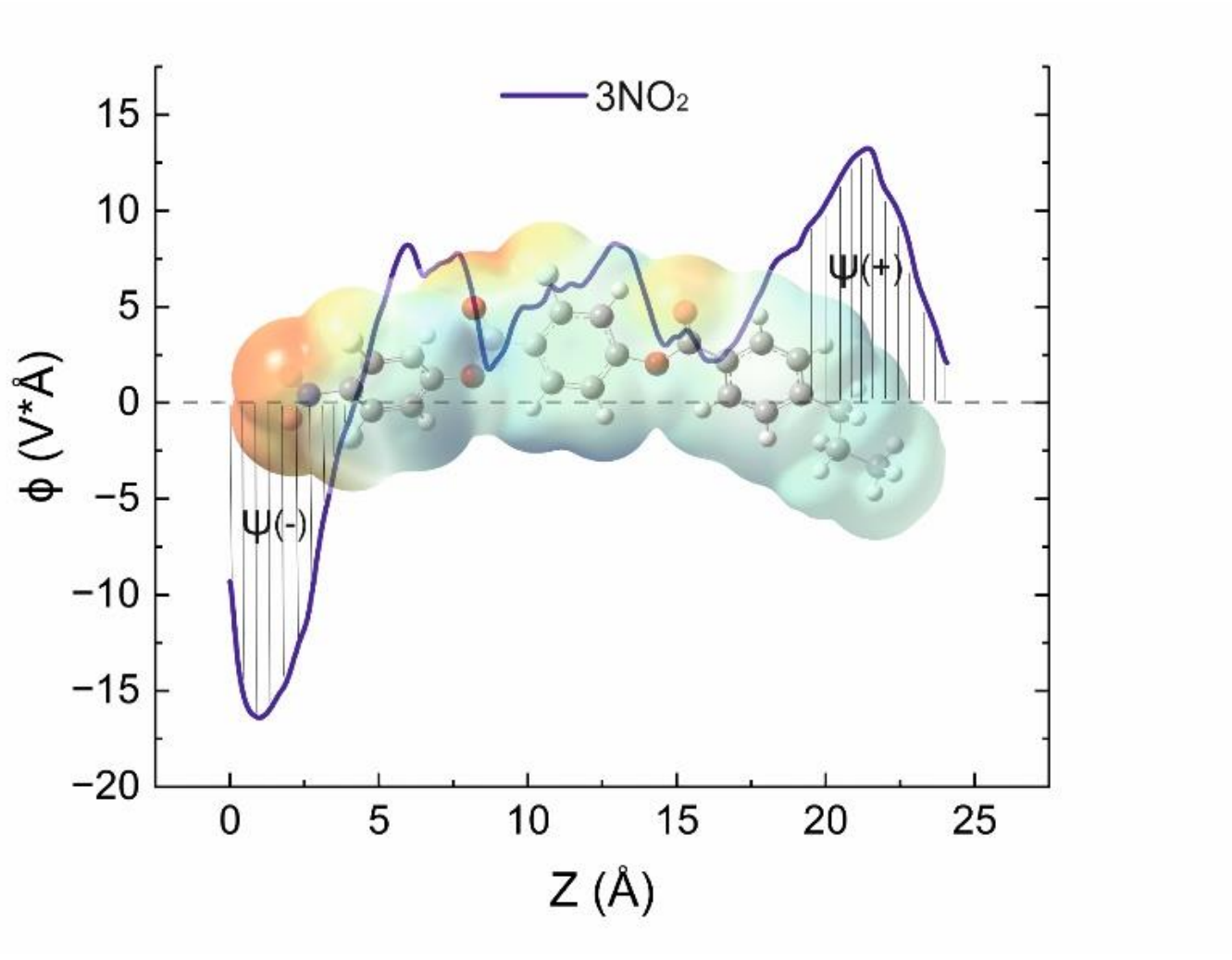

**Figure 6.** 3D-ESP surface and averaged 1D-ESP, scaled by contour length, along the z-axis of the molecule for $3NO_2$ with marked integrated area by vertical lines for $\Psi_{(-)}$ and $\Psi_{(+)}$.

**Table 1.** Dipole moment values, ratios of the temperature ranges of mesophases with polarity on a macro- or microscopic level ($T_{polar}$) to the temperature range of the non-polar mesophase ($T_{non\text{-}polar}$), electronic surface potential (ESP) parameters, and their ratios for $nNO_2$, nCN, and nF analogs.

| Acronym | Dipole moment [Debye] | | | | Polarity of mesophases | | ESP parameters | | |
|---|---|---|---|---|---|---|---|---|---|
| | **$\mu_x$** | **$\mu_y$** | **$\mu_z$** | **$\mu_{total}$** | **ΔTpolar/ ΔTnon-polar in heating** | **ΔTpolar/ ΔTnon-polar in cooling** | **$\Psi_{(-)}$ [VÅ$^2$]** | **$\Psi_{(+)}$ [VÅ$^2$]** | **$\Psi_{(-)}/\Psi_{(+)}$** |
| **$3NO_2$** | 5.34 | 0.33 | 13.49 | 14.51 | N/A[a] | N/A[a] | –45.0 | 44.82 | –1.01 |
| **$4NO_2$** | 5.36 | 0.33 | 13.56 | 14.59 | 0.69 | 2.60 | –45.1 | 49.50 | –0.91 |
| **3CN** | 5.41 | 0.15 | 12.44 | 13.57 | 2.86 | 6.47 | –38.4 | 44.87 | –0.86 |
| **4CN** | 5.42 | 0.15 | 12.52 | 13.64 | 0.57 | 1.47 | –38.4 | 48.13 | –0.80 |
| **3F** | 5.52 | –0.29 | 9.91 | 11.35 | 5.23[b] | 7.21[b] | –25.6 | 42.49 | –0.60 |
| **4F** | 5.53 | –0.29 | 9.99 | 11.43 | 0 | 0 | –25.6 | 45.91 | –0.56 |

a - values were not determined due to the exclusive presence of the $N_F$ polar phase; b - values calculated based on temperature ranges of polar smectic phases.

In this study, the propensity of LC compounds to form polar mesophases is assessed based on the ratio of the temperature ranges of polar to non-polar phases. While the $N_x$ phase is not macroscopically polar, its local antiferroelectric ordering and domain structure with parallel molecular alignment suggest a degree of polarity at the microscopic level. Based on these assumptions, among analogs with identical terminal alkyl chain lengths, the tendency to form polar mesophases follows the increasing polarity of the terminal group (nF < nCN < $nNO_2$;

see Table 1 and Figure 3). An exception arises for the 3F and 3CN analogs, wherein the former exclusively forms polar smectic phases, while the latter stabilizes polar nematic phases. Furthermore, within homologous series featuring the same polar terminal group, an elongation of the terminal alkyl chain diminishes the propensity for polar mesophase formation. To further confirm the macroscopic polarity of the phases in the analyzed compounds, second harmonic generation (SHG) measurements as a function of cooling were conducted for the nX analogs (see Figure S6, ESI). These results confirm the polarity of the $N_F$ phase and its characteristic temperature ranges.

Density functional theory (DFT)-calculated electrical properties provide additional insight into the observed phase behavior. The nCN and $nNO_2$ analogs exhibit comparable dipole moments with similar orientations relative to the molecular long axis (Table 1). In contrast, despite its lower dipole moment, the 3F analog demonstrates a pronounced tendency to form smectic polar phases. This finding corroborates other studies,[49,50] reinforcing that dipole moment magnitude alone is insufficient to predict the emergence of polar mesophases in LC systems. Instead, a more rigorous criterion involves the analysis of the electronic surface potential (ESP) distribution, encompassing both three-dimensional (3D-ESP) and one-dimensional (1D-ESP) averaged profiles along the molecular axis. As demonstrated in recent works,[48–50] this approach aligns with Madhusudana's model, wherein ESP-derived surface charge density distributions play a crucial role in polar phase stabilization. 3D-ESP (Figure S5, ESI) and 1D-ESP (Figure 5) plots reveal characteristic alternating regions of positive and negative potential along the molecular axis for all nF, nCN, and $nNO_2$ analogs. As anticipated, $nNO_2$ compounds exhibit the highest ESP amplitude in the negative electrostatic charged region at one terminal molecular fragment, while the nF analogs display the lowest (Figure 5a, b). Moreover, extending the alkyl chain length leads to a slight broadening and a modest increase in the amplitude of the ESP extrema at the opposite, apolar molecular terminus (Figure 5c, d).

To conduct a more in-depth analysis, the previously described parameters $\Psi_{(-)}$ and $\Psi_{(+)}$ and their ratio were evaluated (Table 1). It was found that a higher $\Psi_{(-)}$ to $\Psi_{(+)}$ ratio correlates with a greater propensity for polar phase formation, both within homologous series and among analogs with the same terminal chain length. Based on this assumption, the 3CN analog would be expected to exhibit a broader range of polar phases relative to non-polar phases than 3F, which has a lower $\Psi_{(-)}$ to $\Psi_{(+)}$ ratio. However, this expectation is not met, likely due to modifications in the ESP distribution along the rigid core upon substitution of F with CN. The enlargement of the minimum in the 1D-ESP at the polar terminal group led to an increase in the maximum within the rigid core near this polar group (Figure 5a and b). As noted in the Introduction, even minor modifications in the ESP distribution along the rigid core can profoundly impact the stability and range of polar phases.[49,50] A substantial reduction in both the amplitude and width of the negative electrostatic charged region at the molecular terminus leads to a decrease in the $\Psi_{(-)}$ parameter, thereby modifying the ESP distribution along the rigid core, as illustrated in Figure 5. This effect, observed in the 3F compound, alongside its noticeably lower dipole moment compared to other analogs, appears to favor the formation of smectic polar phases rather than nematic phases. Moreover, according to the proposed assumption, 3F exhibits a more favorable $\Psi_{(-)}$ to $\Psi_{(+)}$ ratio compared to its longer homolog. The results further suggest that in the case of the 4F homolog, this ratio falls below the threshold required for polar phase formation within the given rigid core structure.

**Phase polarity-terminal groups relationship in other non-chiral LC compounds**

The proposed assumption that a higher ratio of the parameter $\Psi_{(-)}$ to $\Psi_{(+)}$ within the same rigid-core structure favors the stabilization of polar phases might, at first glance, appear to challenge the model developed by Madhusudana.[45] His model postulates that a reduction in charge density at the molecular termini facilitates the emergence of polar phases. However, Madhusudana's calculations primarily considered scenarios in which the amplitudes of surface charge density half-waves at the molecular termini were either equal to each other and comparable to those along the rigid core or lower than the latter. In contrast, the compounds presented in this study, along with numerous other systems exhibiting polar phases,[48,49] typically display the opposite trend. Specifically, the amplitudes of the averaged ESP distribution are highest at the molecular termini and exhibit asymmetry. Madhusudana's work further demonstrated that even a slight increase in the variation of charge density amplitude within the rigid core extends the intermolecular distance range where polar phases are promoted. As system density increases with decreasing temperature, leading to a reduction in intermolecular distance, this variation in amplitude consequently broadens the temperature range over which polar phases can exist. These findings suggest that the primary factor stabilizing parallel molecular arrangements—and thus promoting polarity in liquid crystals—is not merely a reduction in charge density at the molecular termini. Rather, a greater disparity between the charge densities at the termini relative to those along the rigid core, combined with a certain degree of correlation between these distributions, reinforces this effect.

However, stabilization of polar phases is still particularly pronounced when the charge density exhibits a high-frequency sinusoidal variation along the molecular axis.

The proposed hypothesis is supported by numerous recently published studies. Analyzing the findings from various works,[6,49,64] a clear trend emerges: the stability of polar phases increases with the substitution of fluorine atoms on the terminal ring of the rigid core, particularly toward the polar terminal group. As demonstrated by ESP plots along the molecular long axis in this study and others,[49] these fluorine atoms, together with the polar group, generate a surface charge with negative potential at the molecular terminus in Madhusudana's cylindrical model. Consequently, the greater the number of fluorine substitutions in this manner, the higher the ratio of $\Psi_{(-)}$ to $\Psi_{(+)}$. It is also worth noting that Madhusudana suggested that such lateral substitution enhances spatial electrostatic interactions when molecules are sufficiently close together, thereby promoting their parallel alignment.

Empirical evidence also suggests that replacing a terminal fluorine atom with more polar groups such as CN or $NO_2$ enhances the stability of polar phases. For example, in DIO analogs, such modification can lead to the emergence of an enantiotropic $N_X$ phase and an expansion of the $N_F$ phase temperature range.[52] A similar effect is observed in other analogs where 1,3-dioxane units are replaced with ester units: substitution of fluorine with strongly polar CN or $NO_2$ groups extends the ferroelectric phase and induces the appearance of an enantiotropic $N_X$ phase.[53] Additionally, another study [16] reported that in DIO analogs with a laterally substituted nitro group, replacing the terminal fluorine atom with a CN group has been shown to transform the $N_F$ phase from monotropic to enantiotropic while significantly broadening its temperature range.

Furthermore, literature reports indicate that shortening the terminal aliphatic chain enhances polar phase stability. This trend has been documented in DIO analogs,[16] RM734 derivatives,[6,23] and other compounds.[24,65] Such a molecular modification is associated with an increase in the ratio of $\Psi_{(-)}$ to $\Psi_{(+)}$. Analyzing the cited literature, one can also observe that the higher the polarity of the terminal group, the greater the number of homologs exhibiting polar mesophases. However, excessive shortening of the terminal chain may, in some cases, disrupt the molecular weight-to-length ratio, leading to a significant increase in melting temperature and consequently narrowing the range of polar phases.[42]

## Conclusions

In summary, the synthesis and properties of three-ring phenyl esters with a polar terminal nitro group and varying terminal alkyl chain lengths were investigated and characterized. A comprehensive experimental techniques confirmed the presence of an enantiotropic ferroelectric nematic ($N_F$) mesophase in the shorter homolog, whereas the butyl-terminated analog exhibited a sequence of three LC phases, enantiotropic N and $N_x$ phases followed by a monotropic $N_F$ phase. Comparative analysis with previously synthesized analogs bearing different polar terminal groups provided key insights into the role of terminal molecular fragments in governing the formation of polar mesophases in non-chiral LC compounds. DFT calculations were employed to compare the obtained results with Madhusudana's model of polar mesophase formation. These findings suggest that, beyond dipole moment considerations, the spatial modulation of the ESP along the molecular axis plays a crucial role in determining the formation and stabilization of polar mesophases. Moreover, the results confirm and extend conclusions from other studies,[49] demonstrating that a reduction in charge density—particularly of the negative charge—at the molecular termini does not enhance the polarity of the mesophases formed by a given LC compound. This observation directly contradicts Madhusudana's assumption that lower terminal charge density promotes polar phase formation.

Furthermore, a fundamental principle was formulated and validated: within a given homologous series or among analogs with identical terminal chain lengths, a higher ratio of total negative to positive electrostatic charge at the molecular termini correlates with an increased propensity to form polar mesophases. However, when applying this principle, it is essential to consider that even within an identical rigid-core structure, significant changes in the amplitude of extrema in 1D-ESP plots at the molecular termini influence the ESP distribution along the rigid core. This, in turn, determines the polarity of the resulting mesophases. Findings of this work provide deep insights into the molecular origins of polar phases and, consequently, serve as a powerful tool for chemists in predicting new structures of non-chiral LC compounds exhibiting ferro- and antiferroelectric phases.

## Author Contributions

MC prepared bicomponent mixtures, performed optical microscopy and DSC measurements, molecular modeling, analyzed the data, directed the project and wrote the original draft. MM performed spontaneous polarization and dielectric spectroscopy measurements, analyzed the data, and wrote the corresponding parts of the draft. MF recalculated DFT data. DP performed XRD measurements and analyzed the data. NP measured the SHG signal, analyzed the data, and wrote the corresponding parts of the draft. DR prepared the SHG set-up, helped with SHG measurements, and wrote the corresponding methodology part of the draft. MZ performed GC/HPLC–MS analysis. DW synthesized the compounds and wrote the corresponding parts of the draft. All authors contributed to scientific discussions.

**Supporting Information**

Supporting Information is available from the Wiley Online Library or from the author

**Conflicts of interest**

There are no conflicts to declare.

**Acknowledgments**

MCz, MF, MZ and DW acknowledge the financial support from the National Science Centre in Poland (project no. 2024/53/B/ST5/03275). MM thanks the MUT University grant UGB 531-000030-W900-22. NP would like to thank for the financial support the project 24-10247K (Czech Science Foundation). The authors thank Jakub Herman and Mateusz Szala for conducting the NMR spectral measurements. MCz and MF also extend their gratitude to Jordan Hobbs for his valuable advice on generating 1D-ESP plots from DFT data.

## Graphical abstract

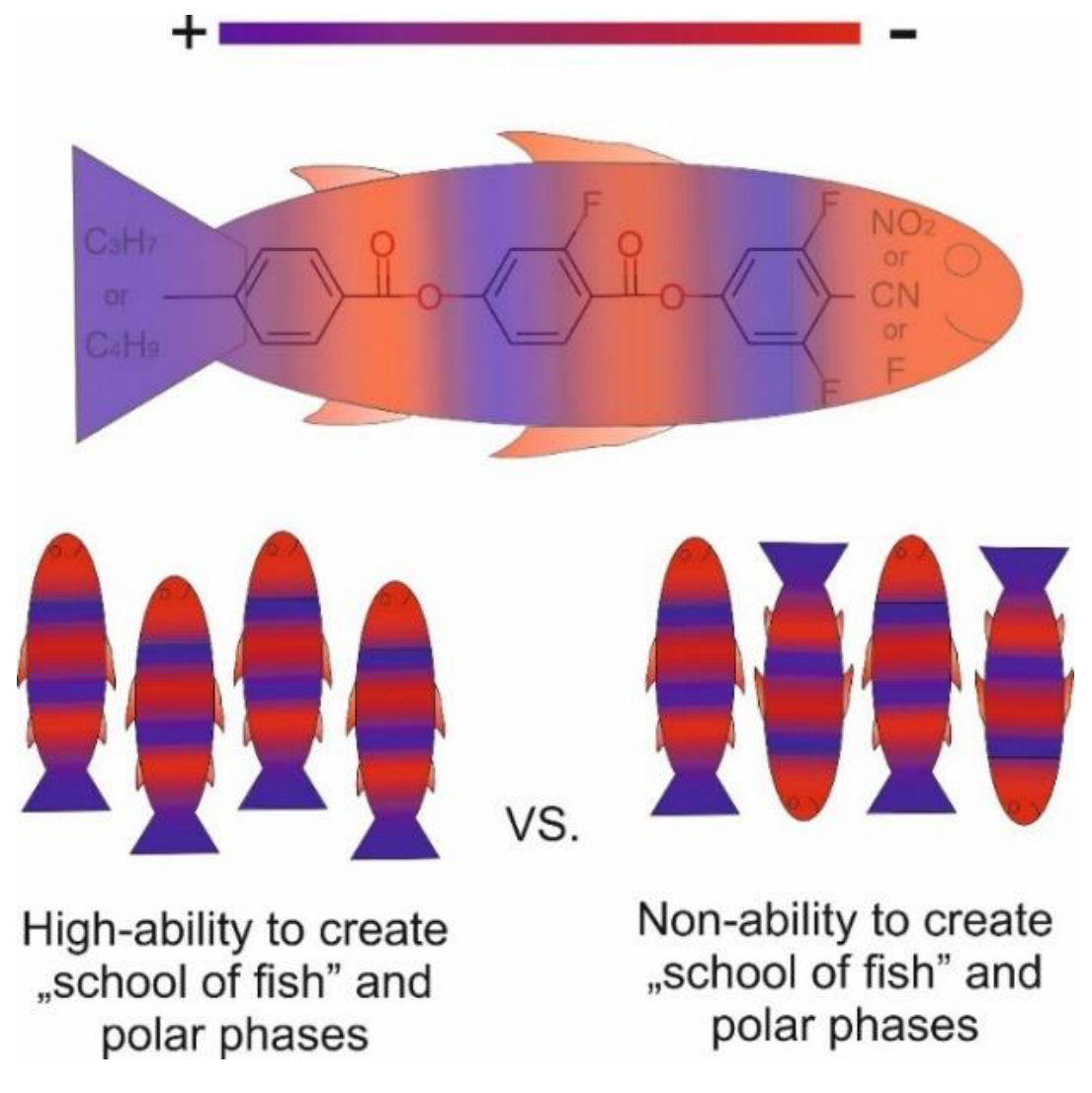

# *Supplementary Material*

**Contents:**

## 1. Experimental methods

### 1.1. Molecular modelling and electronic structure calculations

The dipole moment values for the optimized molecular geometry were calculated using the GAUSSIAN 09 computational chemistry package.[1] Structural optimization and all related calculations were carried out employing the B3LYP hybrid functional combined with the 6-311G+(d,p) basis set, following the methodology established in our previous work.[2] To confirm that the optimized structure corresponds to a true energy minimum, a vibrational frequency analysis was performed, ensuring the absence of imaginary frequencies

Electrostatic potential (ESP) surfaces were computed using the formchk and cubegen utilities of the Gaussian software package. Both electron density and ESP cube files were generated with "fine" data resolution to ensure maximum accuracy. All ESP surfaces presented were visualized at an electron density iso-surface of 0.0004. Data processing was performed using Multiwfn [3] as well as a custom-written script implemented in Python.

To reduce the three-dimensional data to one dimension, for each plane along the z-axis defined in the cube file, an isoline contour was determined where the electron density equals 0.0004. Subsequently, all ESP values along this contour were averaged, based on the assumption that the molecule undergoes free rotation around its principal molecular axis (z-axis).

An additional scaling step was applied by multiplying the obtained values by the contour length, accounting for distortions at the molecular extremities due to decreasing molecular volume. This transformation allows the averaged ESP to be interpreted as the electric flux ($\phi$), representing the strength of the electric field associated with the molecular dipole along the contour. Since the electric flux through a closed surface is equal to the total charge enclosed within that surface, this rescaling can also be understood as a measure of the total charge enclosed within the contour circuit.

In the final step, the electric flux was integrated along the z-axis over a defined molecular segment, enabling the determination of the electric field strength associated with the molecular dipole along the series of contours spanning the z-dimension. This parameter, denoted in the manuscript as $\Psi$ and described by equation (1) in main text of article, can be interpreted as the total electrostatic charge or charge concentration enclosed within the contours forming the iso-surface of the given molecular fragment ($\Psi_{(-)}$ and $\Psi_{(+)}$ for polar and apolar terminal fragment, respectively).

### 1.2.Differential scanning calorimetry

Differential scanning calorimetry (DSC) analyses were conducted using a Netzsch DSC 204 F1 Phoenix calorimeter, calibrated with indium, zinc, and water reference materials. The samples were placed in aluminum crucibles and subjected to a nitrogen atmosphere with a gas flow rate of 20.0 ml/min. Heating and cooling cycles were performed at a rate of 2.0 K/min, and the phase transition temperatures along with the associated enthalpy changes were derived from the corresponding thermograms.

### 1.3. Polarizing optical microscopy

Texture examinations were carried out using an OLYMPUS BX51 polarized optical microscope, coupled with a Linkam TMS93 temperature controller and a THMSE 600 heating stage. The samples were positioned between untreated glass slides to capture their inherent textures.

### 1.4. X-ray diffraction

X-ray diffraction (XRD) studies were conducted using a Bruker GADDS system (microfocus type X-ray tube with Cu anode, 0.5 mm point beam collimator, Vantec 2000 area detector and modified Linkam heating stage). Samples were prepared in a form of one-surface-free droplet on a heated surface.

### 1.5. Optical birefringence

Optical birefringence was measured with a setup composed of a photoelastic modulator (PEM-90, Hinds), a halogen lamp (Hamamatsu LC8) equipped with a narrow bandpass filter (532±3 nm), a photodiode (FLC Electronics PIN-20) and a lock-in amplifier (EG&G 7265). Samples were prepared in glass cells with planar anchoring condition and parallel rubbing assuring uniform alignment of the optical axis in nematic phases.

### 1.6. Second harmonic generation (SHG)

For the temperature-dependent SHG measurements, a Ti:sapphire femtosecond lased amplifier (Spitfire ACE) was used, which consisted of 40-femtosecond pulses with a central wavelength of 800 nm and a pulse repetition rate of 5 kHz. Radiant exposure of individual pulses was set to ~0.01 mJ $cm^{-2}$. The generated and collimated laser beam reached the sample, which was placed into Linkam heating stage equipped with a temperature controller providing an accuracy ±0.1 K. 7 μm HG rubbed in certain direction cells were utilized for the measurements. The rubbing direction and, hence, the long molecular axis was positioned to be parallel to the polarization direction of the incident beam. The measurements were conducted in the transmission mode. The SHG signal generated in the sample was subsequently spectrally filtered with optical dichroic mirrors of the central wavelength 400 nm. The signal was detected with an avalanche photodiode and subsequently amplified with a lock-in amplifier.

### 1.7. Spontaneous electric polarization

The spontaneous polarization was measured using a triangle-wave electric signal. The signal was applied to the 5 μm with bare ITO electrodes by a Hewlett Packard 33120A waveform generator and gained 20 times using an FLC Electronics F20ADI voltage amplifier. The current peak was observed at frequencies of 100 Hz and 5 Hz for $3NO_2$ and $4NO_2$, respectively. The voltage drop across was registered on a Hewlett Packard 54601B oscilloscope. For the $3NO_2$ compound, we used a 1 kΩ resistor, while for $4NO_2$ we used a 10 kΩ resistor. The repolarization current measurement parameters were selected to obtain a single saturated peak in the ferroelectric nematic phase. The temperature was changed and maintained using a Linkam TMS 93 temperature controller in combination with a THMSE 600 heating stage.

### 1.8. Dielectric Spectroscopy

Dielectric spectroscopy studies were performed by using a Hewlett Packard 4294A impedance analyzer in a wide frequency range from 100 Hz to 10 MHz. The measuring field (AC) was 0.1 V. The measurements were done in 5 μm thick cells in the form of parallel plate capacitors. The ITO electrodes were not covered with polyimide alignment layers. The temperature of the cells in the dielectric studies was stabilized using a Linkam TMS 92 temperature controller with a heating stage Linkam THMSE 600, with an accuracy of 0.1 K.

## 2. Chemical Synthesis and Characterization

All chemicals and solvents were used as purchased (J.T. Baker, TM; POCH S.A., Sigma-Aldrich, Avantor Performance Materials Poland S.A, and Merck) and used as received. The purity of intermediates and the main compounds was determined by thin layer chromatography (TLC; with DCM as an eluent and $SiO_2$ as a stationary phase and visualized with 254 or 365 nm UV light) and GC-MS(EI) (Agilent 6890N, Santa Clara, CA, USA). The structures of the final compounds were confirmed by $^1H$, $^{13}C$ NMR, and $^{19}F$ NMR spectroscopy (Bruker, Avance III HD, 500 Hz; $CDCl_3$, Billerica, MA, USA).

### 2.1.Synthetic procedures

The synthetic way of compounds $3NO_2$ and $4NO_2$ was similar to the synthesis of 3F and 4F described in our previous article.[4] The only difference is that 3,4-difluoro-4-nitrophenol (6) is used instead of 2,3,4-trifluorophenol, see Scheme S1.

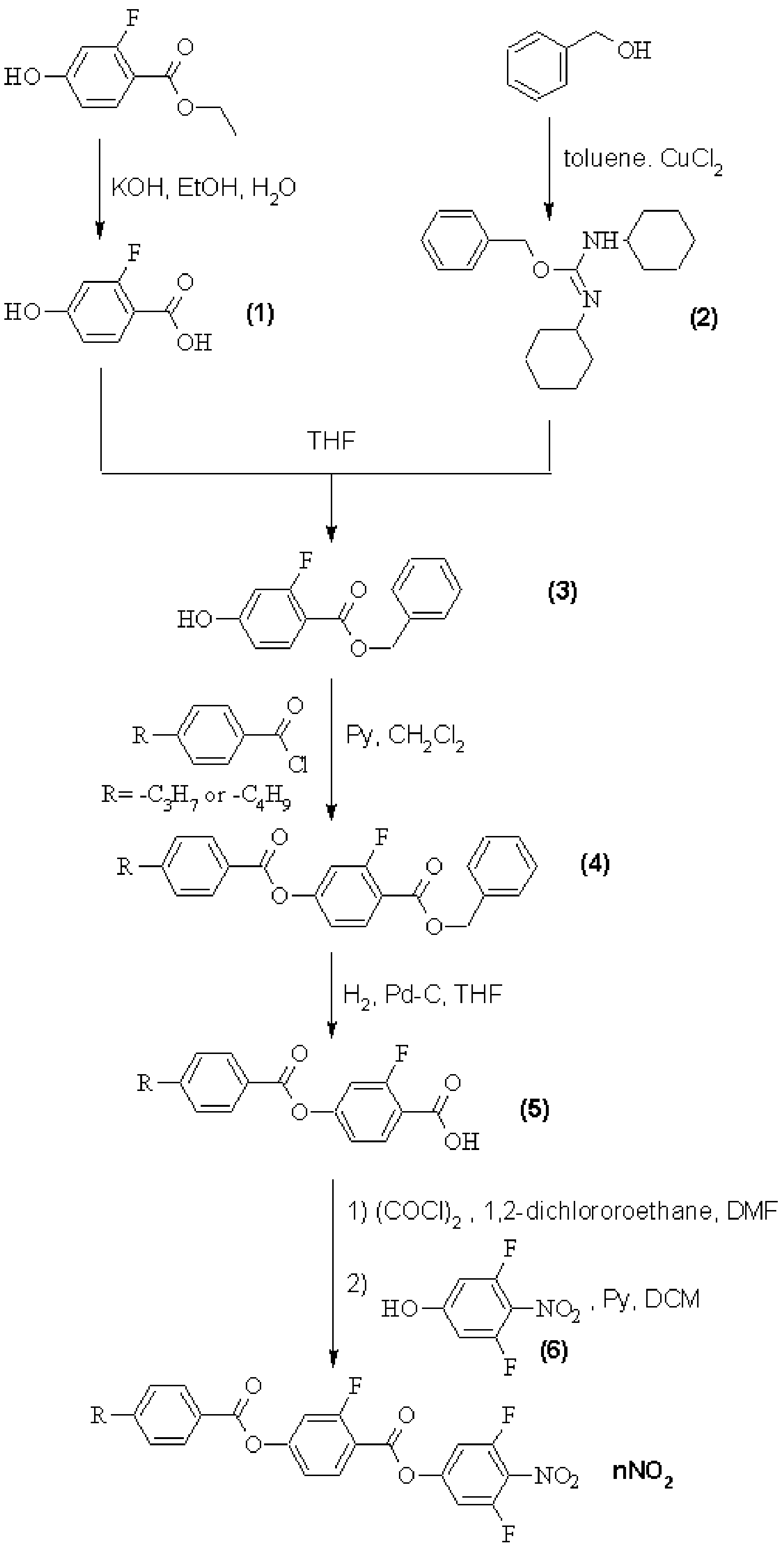

**Scheme S1.** The synthesis pathway of compounds $nNO_2$, R is $C_3H_7$- for $3NO_2$ or $C_4H_9$- for $4NO_2$.

3,4-difluoro-4-nitrophenol **(6)** was obtained by nitration of 3,5-difluorophenol, see Scheme S2.

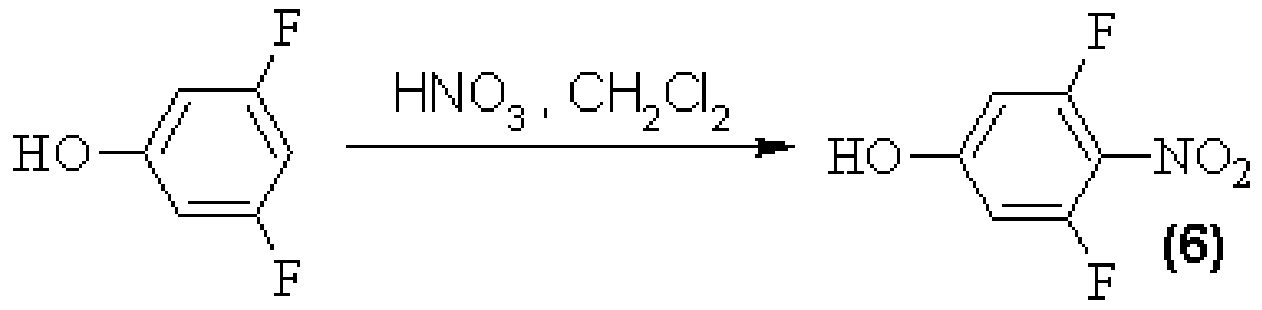

**Scheme S2.** The synthesis pathway of compound **(6)**.

Two isomers obtained during this reaction were separated on a chromatographic column: the desired 3,5-difluoro–4–nitrophenol was a yellow liquid, while 3,5-difluoro–2–nitrophenol was a yellow solid.

## 2.2. Preparative procedure of compound (6)

*3,5-difluoro–4–nitrophenol* **(6)**

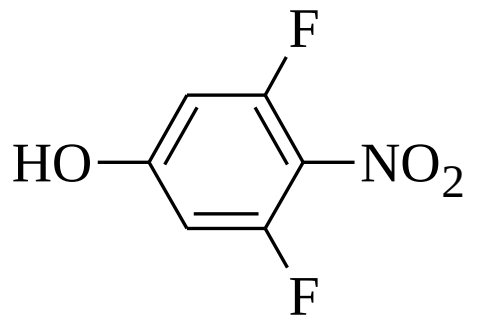

In a three-necked flask placed in an ice/water bath, equipped with a mechanical stirrer, a thermometer and a dropping funnel, 3,5-difluorophenol (14.1 g; 0.11 mol) and dichloromethane (150 ml) were placed. The reaction mixture was stirred at 0 °C, and then fuming nitric acid (15 ml) was added dropwise, maintaining the temperature of 0 °C for two hours. Then, cold water was added, and the phases were separated. The water phase was extracted with dichloromethane (2×50 ml). The organic layer was washed three times with brine (2×50 ml), dried over $MgSO_4$, and concentrated at low pressure. The crude product was purified by chromatography column ($SiO_2$; ethyl acetate/hexane in a volume ratio 1/9). Yield: 5.8 g (30%). GC: 95.6 %; MS(EI) m/z: 176 $[M+H]^+$.

## 2.3. Characterization of final compounds

*3,5-difluoro-4-nitrophenyl 2-fluoro-4-[(4-propylbenzoyl)oxy]benzoate* **$3NO_2$**

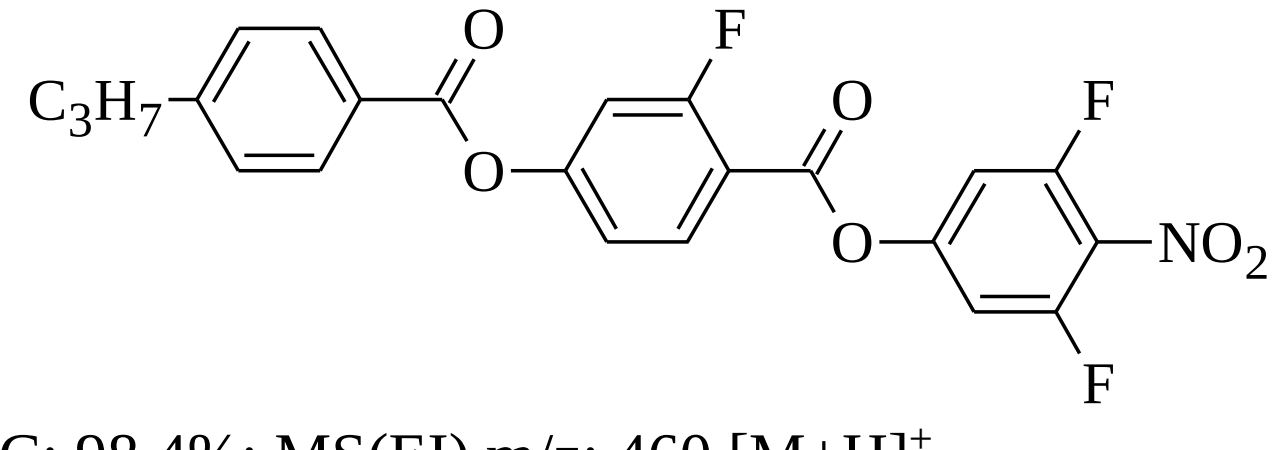

Yield: 0,7 g (70%); GC: 98.4%; MS(EI) m/z: 460 $[M+H]^+$.

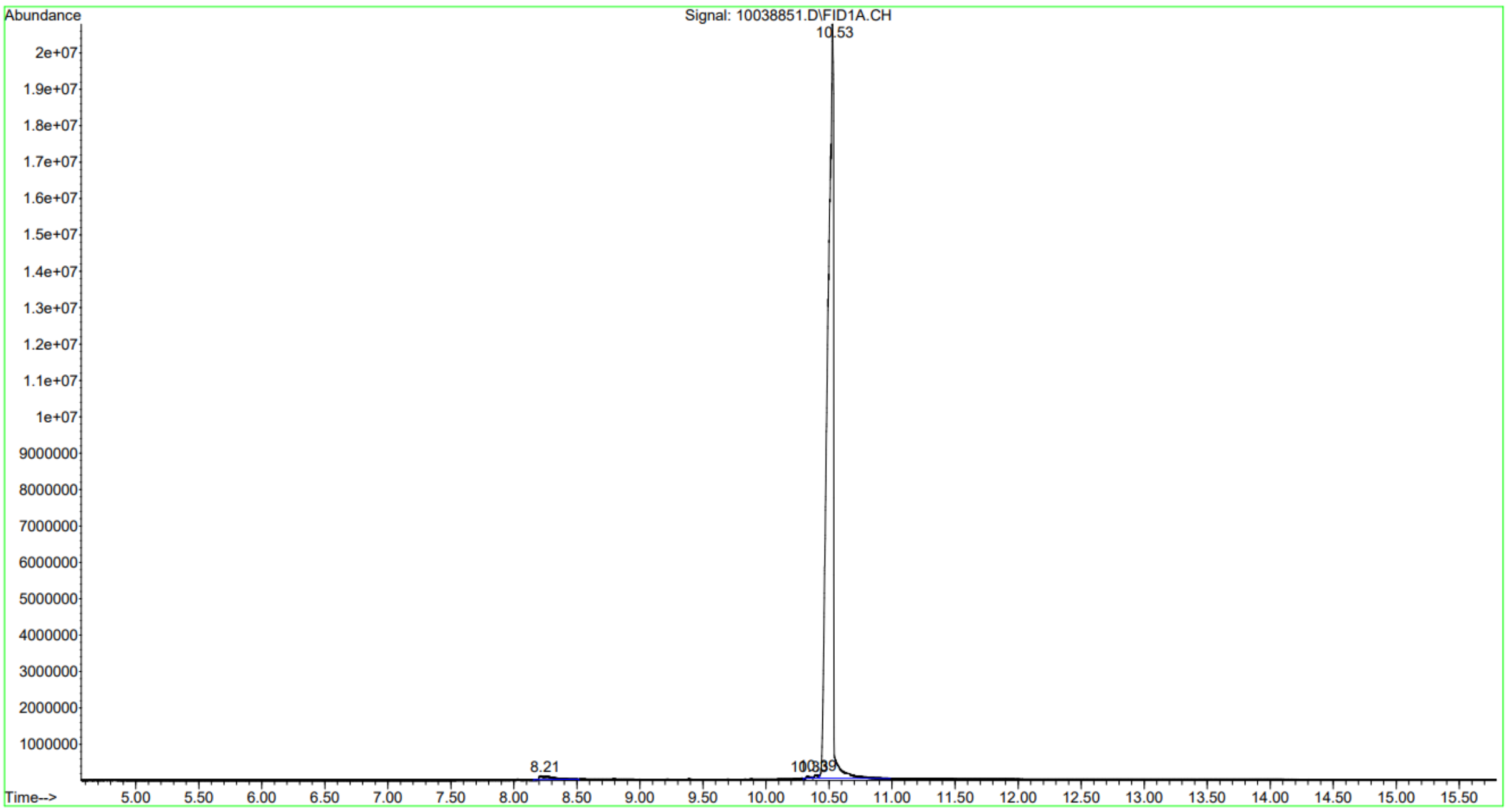

| peak # | R.T. min | Start min | End min | PK TY | peak height | corr. area | corr. % max. | % of total |
|---|---|---|---|---|---|---|---|---|
| 1 | 8.214 | 8.157 | 8.515 | M | 109196 | 8040204 | 1.16% | 1.145% |
| 2 | 10.327 | 10.296 | 10.366 | M | 73132 | 1440330 | 0.21% | 0.205% |
| 3 | 10.392 | 10.370 | 10.422 | M | 99542 | 2123493 | 0.31% | 0.302% |
| 4 | 10.529 | 10.422 | 10.985 | M | 21105805 | 690580700 | 100.00% | 98.347% |

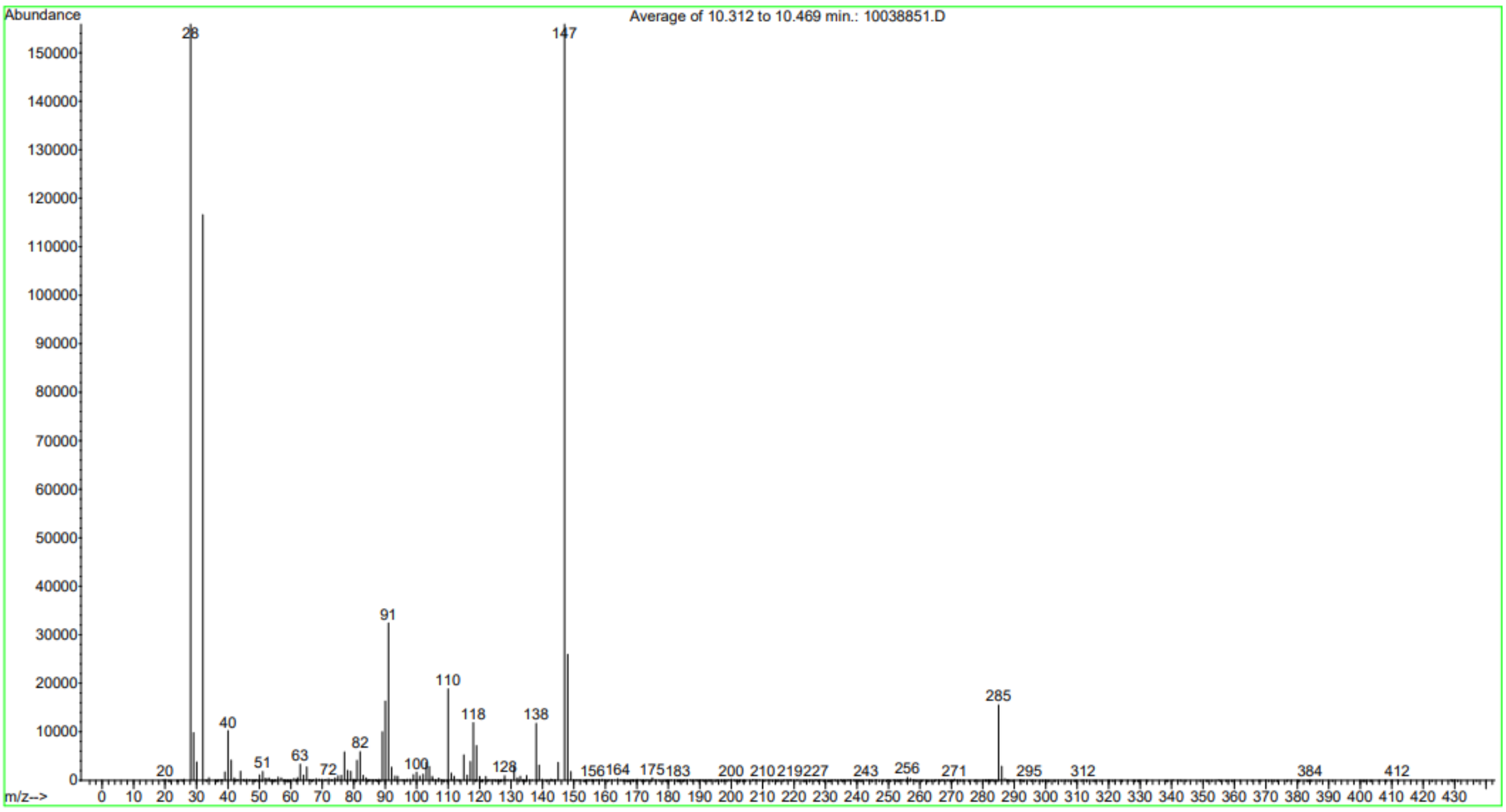

[1]H NMR (500 MHz, $CDCl_3$) δ/ppm: 1.0 (t, *J*=7.3 Hz, 3H, $-CH_3$), 1.7 (td, *J*=15.0, 7.3 Hz, 2H, $-CH_2-$), 2.7 (m, 2H, $-CH_2-$), 7.2 (dd, *J*=12.9, 4.6 Hz, 2H, Ar-H), 7.25 (m, 2H, Ar-H), 7.4 (d, *J*=8.2 Hz, 2H, Ar-H), 8.1 (d, J=8.2 Hz, 2H, Ar-H), 8.2 (s, 1H, Ar-H).

$^{13}C$ NMR (126 MHz, $CDCl_3$) δ/ppm: 13.8 (s, 1C), 24.2 (s, 1C), 38.2 (s, 1C), 107.4 (s, 1C), 107.6 (s, 1C), 111.5 (s, 1C), 111.7 (s, 1C), 113.7 (d, *J*=9.0 Hz, 1C), 118.3 (d, *J*=3.6 Hz, 1C), 125.8 (s, 1C), 129.0 (s, 2C), 130.5 (s, 2C), 133.6 (s, 1C), 150.1 (s, 1C), 153.2 (t, *J*=13.6 Hz, 1C), 154.2 (d, *J*=3.6, 1C), 156.3 (d, *J*=3.6 Hz, 1C), 157.1 (s, 1C), 160.4 (d, *J*=4.5 Hz, 1C), 162.0 (s, 1C), 164.0 (s, 1C).
$^{19}F$ NMR (470 MHz, $CDCl_3$) δ/ppm: -103.3 (t, *J*=9.2 Hz, 1F), -116.0 (d, *J*=8.9 Hz, 2F).

*3,5-difluoro-4-nitrophenyl 2-fluoro-4-[(4-butylbenzoyl)oxy]benzoate* **$4NO_2$**

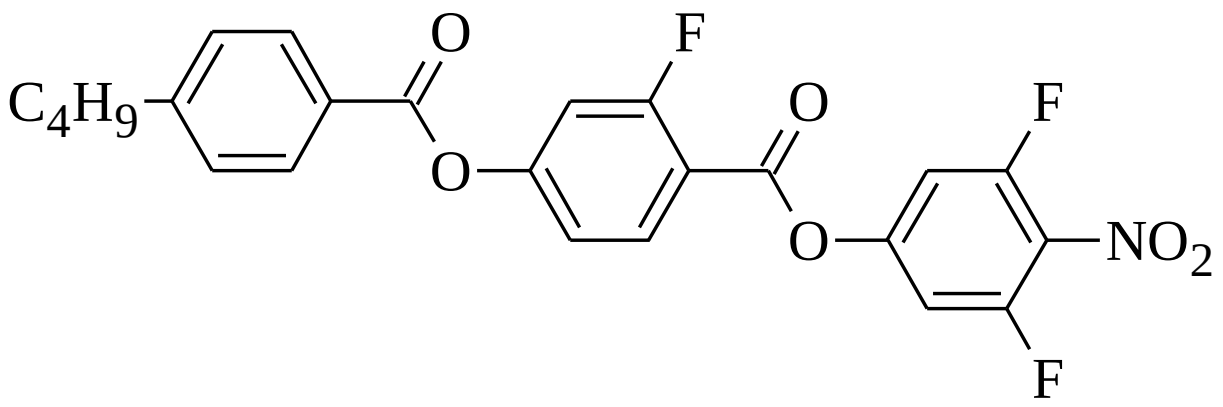

Yield: 0,7 g (70%); GC: 98.3%; MS(EI) m/z: 474 $[M+H]^+$.

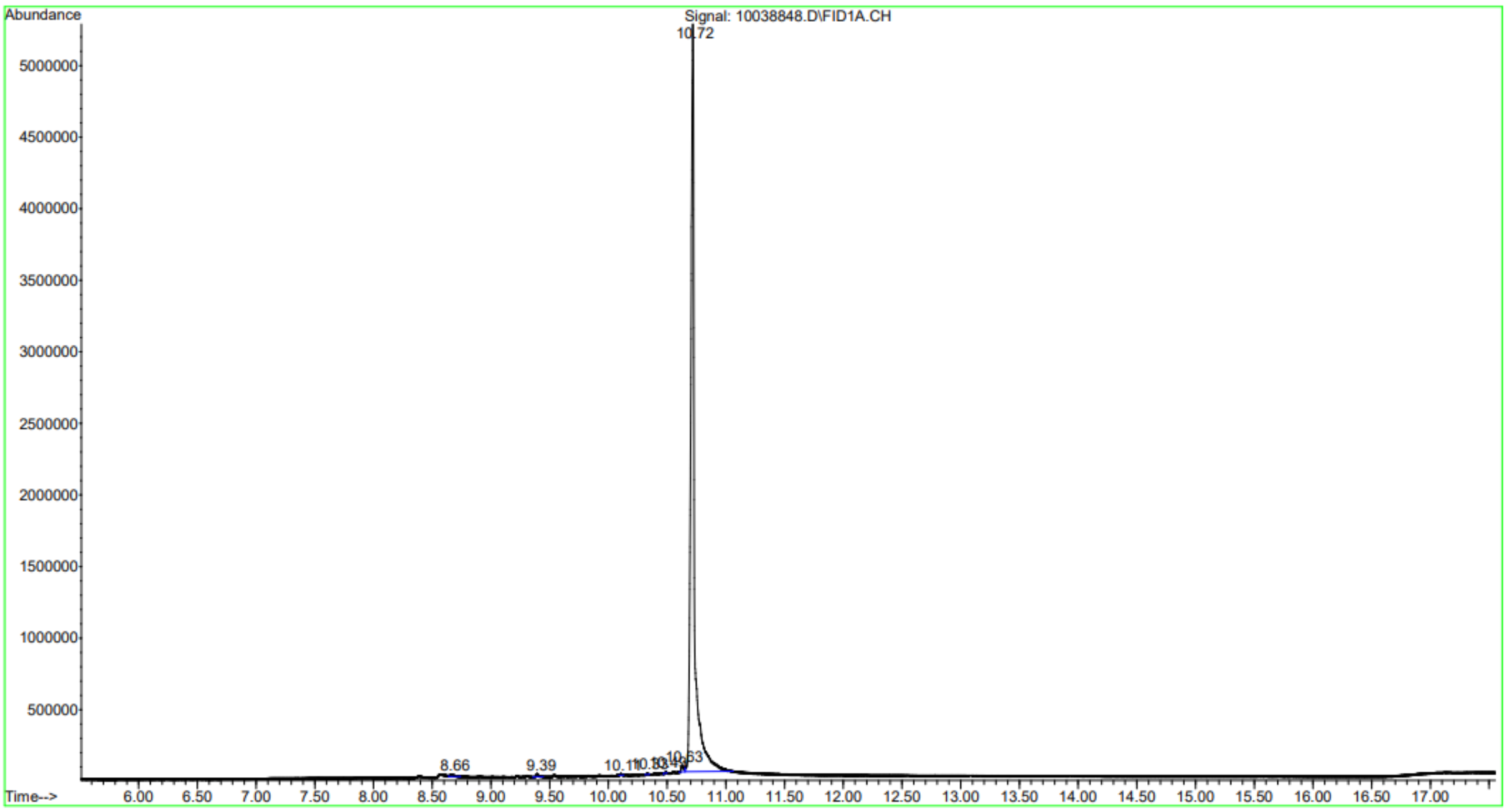

| peak # | R.T. min | Start min | End min | PK TY | peak height | corr. area | corr. % max. | % of total |
|---|---|---|---|---|---|---|---|---|
| 1 | 8.657 | 8.639 | 8.742 | M | 17099 | 540334 | 0.46% | 0.455% |
| 2 | 9.393 | 9.362 | 9.445 | M | 22114 | 404249 | 0.35% | 0.341% |
| 3 | 10.106 | 10.090 | 10.128 | M | 12189 | 143415 | 0.12% | 0.121% |
| 4 | 10.334 | 10.314 | 10.350 | M | 9023 | 110045 | 0.09% | 0.093% |
| 5 | 10.491 | 10.469 | 10.498 | M | 9627 | 78355 | 0.07% | 0.066% |
| 6 | 10.627 | 10.603 | 10.653 | M | 49754 | 762961 | 0.65% | 0.643% |
| 7 | 10.721 | 10.654 | 11.060 | M | 5569019 | 116638589 | 100.00% | 98.282% |

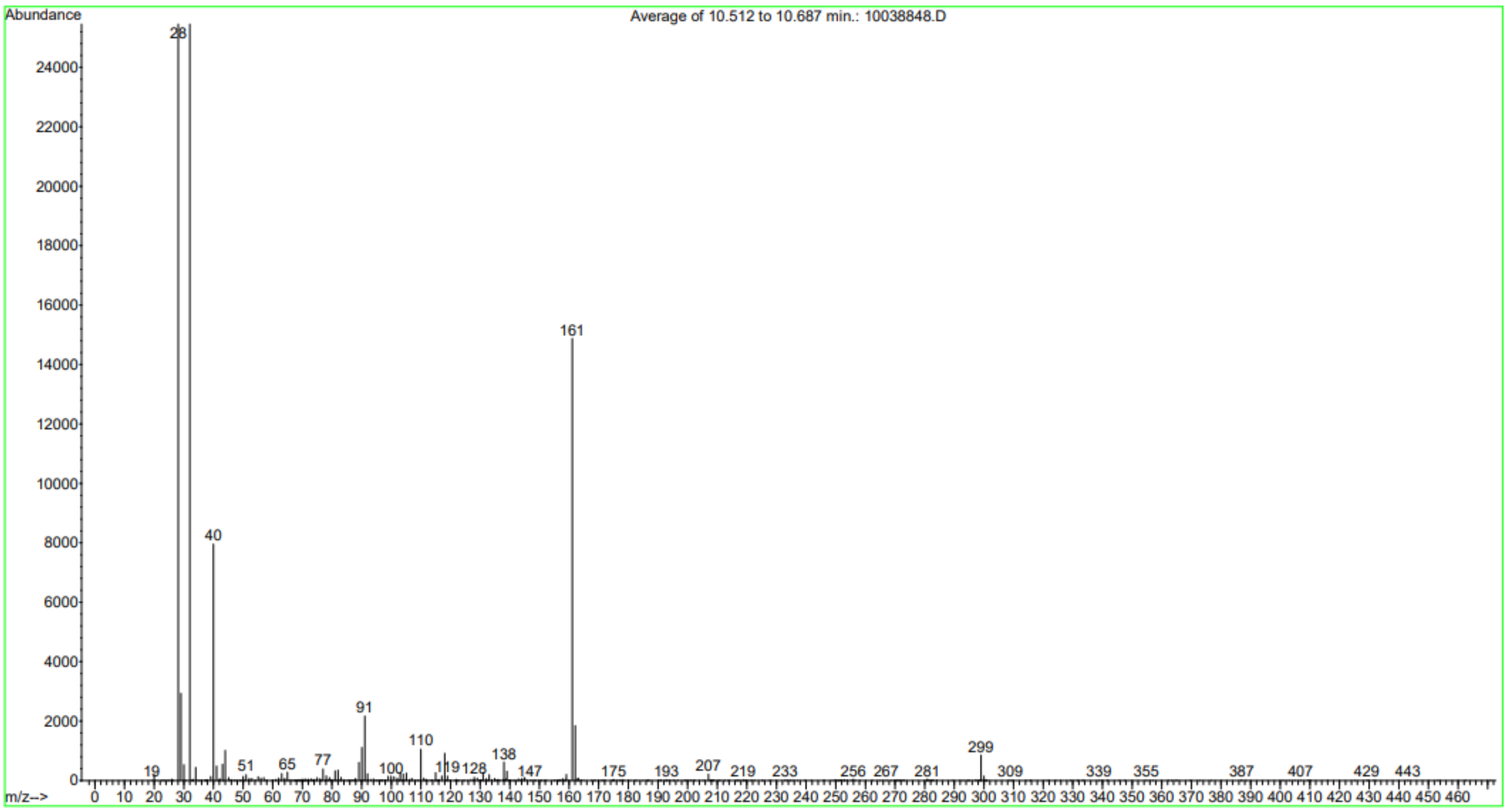

[1]H NMR (500 MHz, $CDCl_3$) δ/ppm: 1.0 (t, *J*=7.3 Hz, 3H, $-CH_3$), 1.4 (td, *J*=14.9, 7.5 Hz, 2H, $-CH_2-$), 1.7 (m, 2H, $-CH_2-$), 2.7 (t, *J*=8 Hz, 2H, $-CH_2-$), 7.2 (dd, *J*=12.9, 4.5 Hz, 2H, Ar-H), 7.25 (m, 2H, Ar-H), 7.4 (d, *J*=8.2 Hz, 2H, Ar-H), 8.1 (d, J=8.2 Hz, 2H, Ar-H), 8.2 (s, 1H, Ar-H).

[13]C NMR (126 MHz, $CDCl_3$) δ/ppm: 13.9 (s, 1C), 22.4 (s, 1C), 33.2 (s, 1C), 35.9 (s, 1C), 107.4 (s, 1C), 107.6 (s, 1C), 111.5 (s, 1C), 111.7 (s, 1C), 113.7 (d, *J*=10.0 Hz, 1C), 118.3 (d, *J*=3.6 Hz, 1C), 125.8 (s, 1C), 129.0 (s, 2C), 130.5 (s, 2C), 133.6 (s, 1C), 150.4 (s, 1C), 153.2 (t, *J*=12.7 Hz, 1C), 154.2 (d, *J*=3.6, 1C), 156.3 (d, *J*=3.6 Hz, 1C), 157.0 (s, 1C), 160.4 (d, *J*=4.5 Hz, 1C), 162.0 (s, 1C), 164.0 (s, 1C).

[19]F NMR (470 MHz, $CDCl_3$) δ/ppm: -103.3 (t, *J*=8.5 Hz, 1F), -116.0 (d, *J*=9.6 Hz, 2F).

## 3. Supplementary results

**Table S1.** The phase transition temperatures [°C] (onset point), in bold font, and corresponding enthalpy changes [kJ mol[-1]], in italic font, as well as the scaled entropy change, ΔS/R, of the members of nX analogs, from DSC measurements determined during heating (upper rows) and cooling (down rows); values given in brackets were determined for monotropic phase.

| Acronym | Cr | T[°C]<br>*ΔH[kJmol$^{-1}$]*<br>ΔS /R[a.u] | $N_F$ | T[°C]<br>*ΔH[kJmol$^{-1}$]*<br>ΔS /R[a.u.] | $N_X$ | T[°C]<br>*ΔH[kJmol$^{-1}$]*<br>ΔS /R[a.u.] | N | T[°C]<br>*ΔH[kJmol$^{-1}$]*<br>ΔS /R[a.u.] | Iso |
|---|---|---|---|---|---|---|---|---|---|
| **3NO$_2$** | • | **128.2**<br>*21.47*<br>6.43<br>**102.4**<br>*-19.00*<br>-6.08 | • | | - | | - | **147.6**<br>*3.89*<br>1.11<br>**149.9**<br>*-8.70*<br>-2.47 | • |
| **4NO$_2$** | • | **115.9**<br>*22.03*<br>6.81<br>**92.0**<br>*-22.73*<br>-7.48 | (•) | -<br>**(114.9)**<br>*(-0.62)*<br>-0.19 | • | **124.5**<br>*0.04*<br>0.012<br>**123.8**<br>*-0.04*<br>-0.012 | • | **137.0**<br>*0.43*<br>0.13<br>**136.0**<br>*-0.44*<br>-0.13 | • |
| **3CN []** | • | **109.6**<br>*25.28*<br>7.94<br>**39.3**<br>*-9.94*<br>-3.82 | • | **156.2**<br>*0.59*<br>0.16<br>**157.7**<br>*-0.56*<br>-0.15 | • | **168.9**<br>*0.03*<br>0.008<br>**173.9**<br>*-0.05*<br>-0.013 | • | **189.5**<br>*0.76*<br>0.19<br>**194.7**<br>*-0.66*<br>-0.17 | • |
| **4CN []** | • | **115.0**<br>*35.76*<br>11.08<br>**80.2**<br>*-23.89*<br>-8.13 | (•) | -<br>**(102.8)**<br>*(-0.39)*<br>-0.12 | • | **137.1**<br>*0.02*<br>0.006<br>**137.3**<br>*-0.02*<br>-0.006 | • | **175.7**<br>*0.56*<br>0.15<br>**176.9**<br>*-0.67*<br>-0.18 | • |

| Acronym | Cr | T[°C]<br>*ΔH[kJmol$^{-1}$]*<br>ΔS /R[a.u] | $SmC_F$ | T[°C]<br>*ΔH[kJmol$^{-1}$]*<br>ΔS /R[a.u] | $SmA_F$ | T[°C]<br>*ΔH[kJmol$^{-1}$]*<br>ΔS /R[a.u] | N | T[°C]<br>*ΔH[kJmol$^{-1}$]*<br>ΔS /R[a.u] | Iso |
|---|---|---|---|---|---|---|---|---|---|
| **3F []** | • | **104.1**<br>*20.92*<br>6.67<br>**89.4**<br>*-10.72*<br>-3.55 | (•) | -<br>**(99.1)**<br>*(-0.03)*<br>-0.01 | • | **131.8**<br>*4.44*<br>1.318<br>**130.5**<br>*-4.40*<br>-1.311 | • | **137.1**<br>*0.43*<br>0.13<br>**136.2**<br>*-0.39*<br>-0.12 | • |

| Acronym | Cr | T[°C]<br>*ΔH[kJmol$^{-1}$]*<br>ΔS /R[a.u] | $SmC_M$ | T[°C]<br>*ΔH[kJmol$^{-1}$]*<br>ΔS /R[a.u] | SmA | T[°C]<br>*ΔH[kJmol$^{-1}$]*<br>ΔS /R[a.u] | N | T[°C]<br>*ΔH[kJmol$^{-1}$]*<br>ΔS /R[a.u] | Iso |
|---|---|---|---|---|---|---|---|---|---|
| **4F []** | • | **101.6**<br>*26.88*<br>8.62<br>**84.2**<br>*-14.85*<br>-5.00 | (•) | -<br>**(92.0)**<br>*(-0.12)*<br>-0.04 | • | **124.7**<br>*2.45*<br>0.74<br>**123.2**<br>*-2.26*<br>-0.68 | • | **130.5**<br>*0.46*<br>0.14<br>**129.5**<br>*-0.48*<br>-0.14 | • |

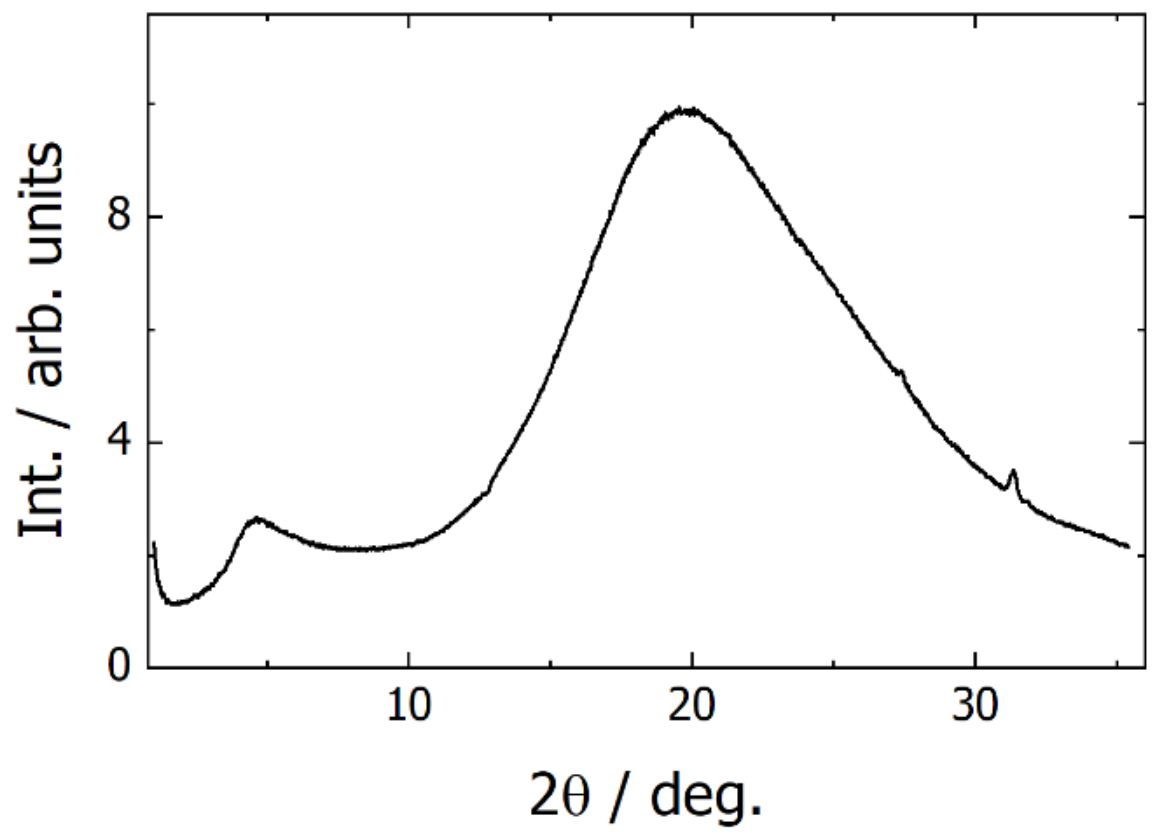

**Figure S1.** Signal intensity vs. diffraction angle obtained by integration of 2D pattern over azimuthal angle. The weak, sharp signal visible at 2θ = 31 deg. Is an artefact due to the metal sample holder.

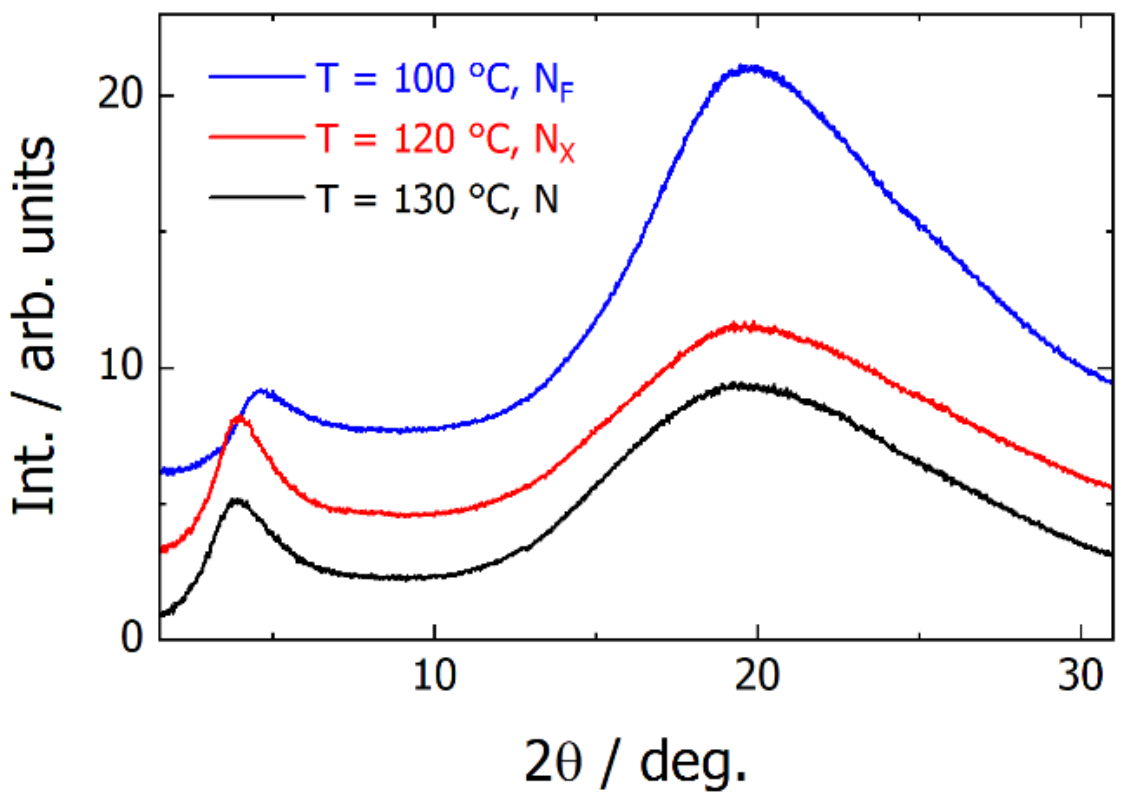

**Figure S2.** Signal intensity vs. diffraction angle obtained by integration of 2D patterns over azimuthal angle, curves are vertically shifted for clarity of presentation. *Note that the change of relative low- and high-angle signal intensities in* $N_F$ *phase results from the change of the sample alignment, and does not reflect changes in phase structure.*

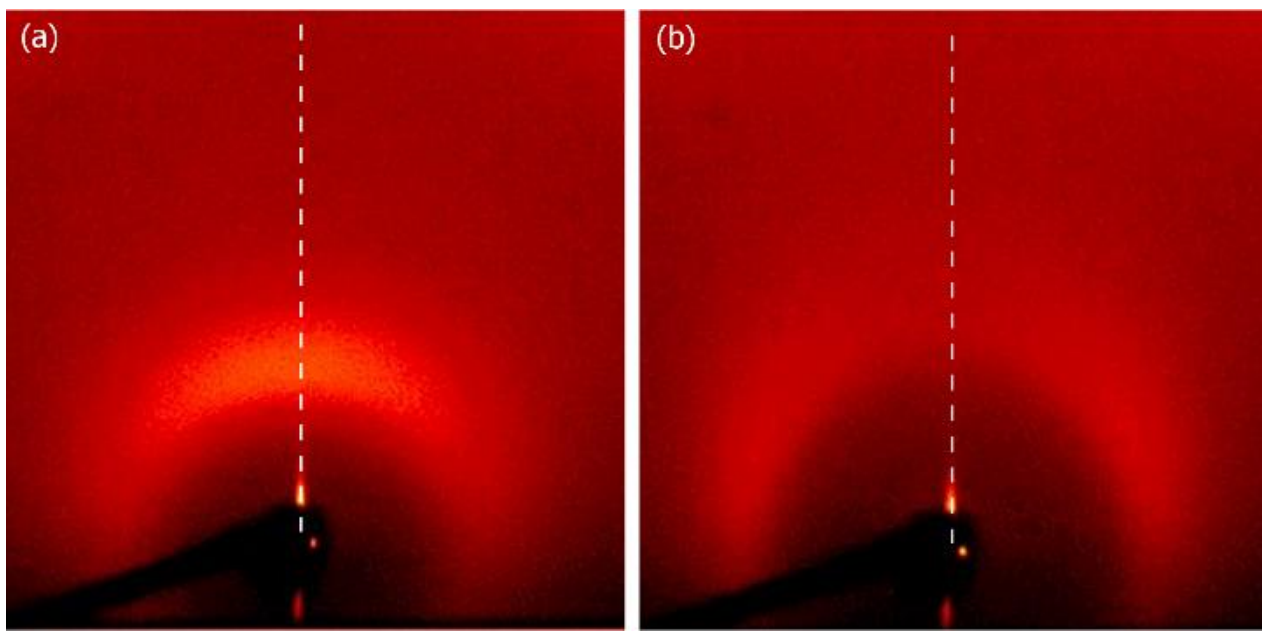

**Figure S3.** Small angle part of 2D XRD patterns recorded for $4NO_2$ in (a) N and (b) $N_F$ phases. In paraelectric nematic phase diffraction signal is positioned along the sample surface normal (dashed white line) evidencing preferable homeotropic alignment in this phase in one-surface-free sample. In ferroelectric NF phase the preferred orientation of director (and thus electric polarization) changes to planar in order to avoid surface charges, and the maximum of diffraction signal moves to horizontal position.

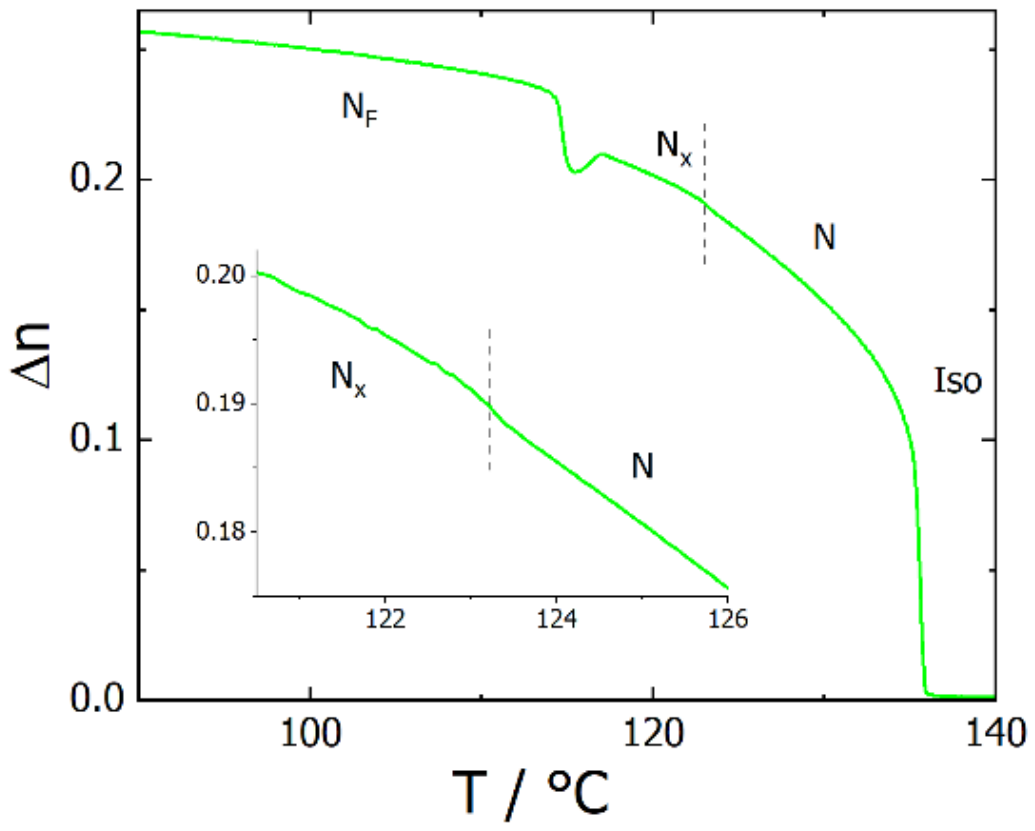

**Figure S4.** Optical birefringence as a function of temperature for $4NO_2$ compound. Measurements were performed with green light ($\lambda$=532 nm) for material placed in 1.5-µm-thick cell with planar anchoring condition.

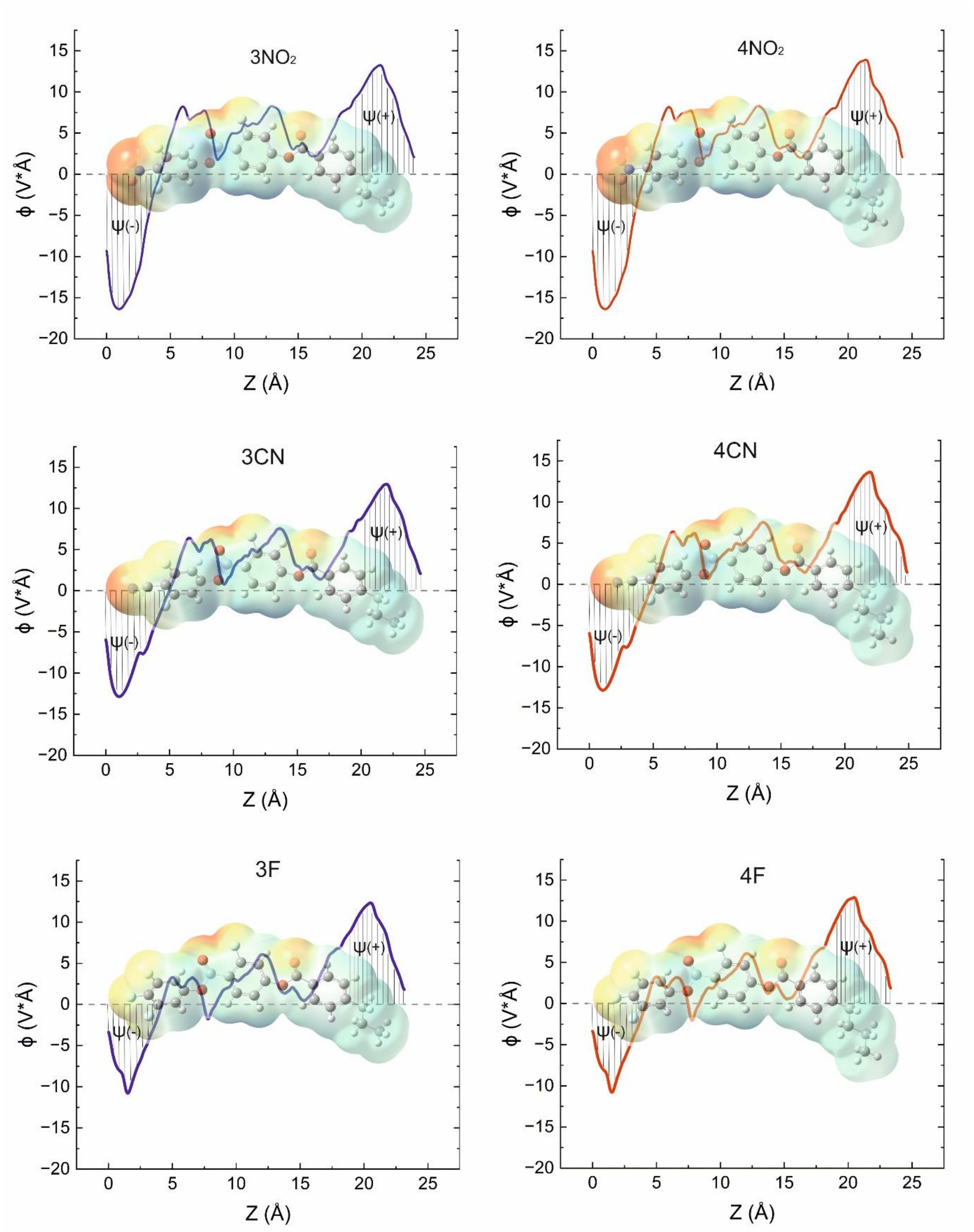

**Figure S5.** 3D-ESP surface and averaged 1D-ESP, scaled by contour length, along the z-axis of the molecule at an electron density isosurface of 0.0004 calculated at the B3LYP-6-311G+(d,p) level of DFT for analogs nX; vertical lines show integrated area for $\Psi_{(-)}$ and $\Psi_{(+)}$.

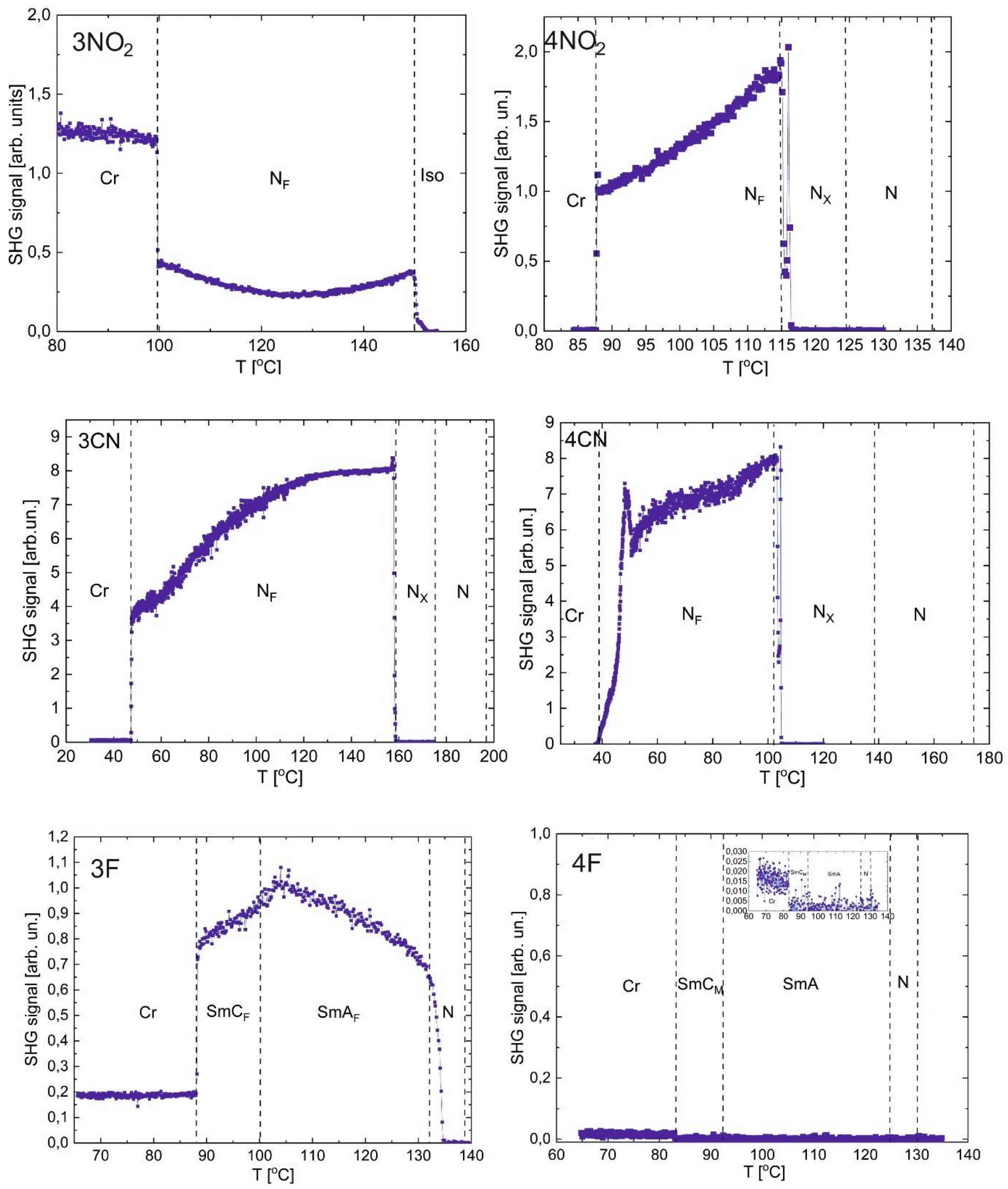

**Figure S6.** SHG signal as a function of cooling for analogs nX.

## 4. Supplemental references